\documentclass[sigconf]{acmart}
\usepackage{tikz}
\usepackage{amsmath}
\usepackage{filecontents}
\usepackage{subcaption}
\usepackage[toc,page]{appendix}
\usepackage{cleveref}
\crefformat{section}{\S#2#1#3} 
\crefformat{subsection}{\S#2#1#3}
\crefformat{subsubsection}{\S#2#1#3}

\acmConference[MemSys '20]{MemSys '20: The International Symposium on Memory
Systems}{ Sept 2020}{Washington, DC}
\acmPrice{15.00}
\acmISBN{978-1-4503-XXXX-X/18/06}

\begin{document}

\title{Efficient Orchestration of Host and Remote Shared Memory
\\for Memory Intensive Workloads}


\newcommand{\tsc}[1]{\textsuperscript{#1}} 
\author{Juhyun Bae$^1$, Gong Su$^2$, Arun Iyengar$^2$, Yanzhao Wu$^1$, Ling Liu$^1$}
\affiliation{
\institution{\vskip .2cm}
\institution{$^1$Georgia Institute of Technology, Atlanta, GA30329, USA}
\institution{$^2$IBM Thomas J. Watson Research, New York, USA}
}


\begin{abstract}
Since very few contributions to the development of an unified memory orchestration framework for efficient management of both host and remote idle memory have been made, we present Valet, an efficient approach to orchestration of host and remote shared memory for improving performance of memory intensive workloads. The paper makes three original contributions. First, we redesign the data flow in the critical path by introducing a host-coordinated memory pool that works as a local cache to reduce the latency in the critical path of the host and remote memory orchestration. Second, Valet utilizes unused local memory across containers by managing local memory via Valet host-coordinated memory pool, which allows containers to dynamically expand and shrink their memory allocations according to the workload demands. Third, Valet provides an efficient remote memory reclaiming technique on remote peers, based on two optimizations: (1) an activity-based victim selection scheme to allow the least-active-block of data to be selected for serving the eviction requests and (2) a migration protocol to move the least-active-block of data to less-memory-pressured remote node. As a result, Valet can effectively reduce the performance impact and migration overhead on local nodes. Our extensive experiments on both NoSQL systems and Machine Learning (ML) workloads show that Valet outperforms existing representative remote paging systems by up to 282$\times$ throughput improvement and up to 98\% latency decrease over conventional OS swap facility for big data and ML workloads, and by up to 5.3$\times$ throughput improvement and up to 80\% latency decrease over the state-of-the-art remote paging systems. Valet is open sourced at https://github.com/git-disl/Valet.
\end{abstract}


\maketitle

\section{Introduction}

Data-intensive and latency-demanding applications~\cite{Memcached, Redis, VoltDB, Hadoop, Spark} are typically deployed using the application deployment models, comprised of containers, virtual machines (VMs), and/or executors/JVMs.
These applications enjoy high throughput and low latency if they are served entirely from memory. Challenges on these applications increase as workload size becomes larger. When these applications cannot fit their working sets in physical memory of their containers/VMs/executors, they suffer large performance loss in latency, throughput and completion time due to excessive page faults and thrashing.

Most of the existing research studied the above problems and proposed to increase effective memory capacity of VMs/containers by leveraging remote idle memory resources. These proposals promote new architectures and new hardware design for memory disaggregation~\cite{hpthememory,intel-RSA,lim2009disaggregated,Lim+HPCA2012,gao2016network, aguilera2017remote}, or new programming models~\cite{Nelson+-usenixATC2015,PowerPiccolo-OSDI2010}. But they lack of desired transparency at OS, network stack, or application level, hindering their practical applicability. Other efforts~\cite{Evangelos, Tia, Shuang, Haogan, Umesh, Juncheng, Hikari} promotes remote paging with transparency to improve OS paging performance by exploiting the disk-network latency gap via unused remote memory~\cite{nbdX, AndersonNeefe-1994, Chen+workshop2008, Dwarkadas+cashmere-VLM-IPPS1999, FlourisMarkatos-JCC-1999, Juncheng, LiangNoronhaPanda-CC2005, MarkatosDramitinos-usenixATC1996, nbdX, Tia,zhang2015hybridswap,hao2016tail,Spongefiles}. However, most existing solutions~\cite{nbdX, lim2009disaggregated, Juncheng, Tia, zhang2015hybridswap} suffer from high latency limitations due to remote node memory allocation overhead due to receiver-side CPU involvement and the scale-out performance with the large workload. Moreover, existing research efforts have been dedicated either to consolidation of host idle memory across VMs/Containers on the same host or focused on remote memory disaggregation. Very few contributes to the development of an unified memory orchestration framework for efficient management of both host and remote idle memory. 

In this paper we presents Valet, an efficient orchestration of host and remote shared memory for big data and machine learning workloads that are memory-intensive in nature. Valet by design aims to address the following three common problems inherent in existing remote memory systems. First, they have latency overhead in the performance critical path due to dynamic connection setup to the remote node(s) and remote memory mapping or disk access scenarios~(\cref{latecnyoverheadproblem}). Second, recent effort \cite{Juncheng} shows the benefit of remote memory paging with RDMA network and the limitation due to eviction impact when remote node evicts data of local nodes~(\cref{evictionimpactproblem}). Finally, with the increasing popularity of Container as a Service (CaaS)~\cite{ContainerMarket}, the container-wide memory imbalance~(\cref{containerimbalanceproblem}) involves managing both node-level memory imbalance and cluster-wide memory imbalance, which pose non-trivial technical challenges~\cite{Ling}. 

\begin{figure}[t]
\centering
\includegraphics[width=80mm]{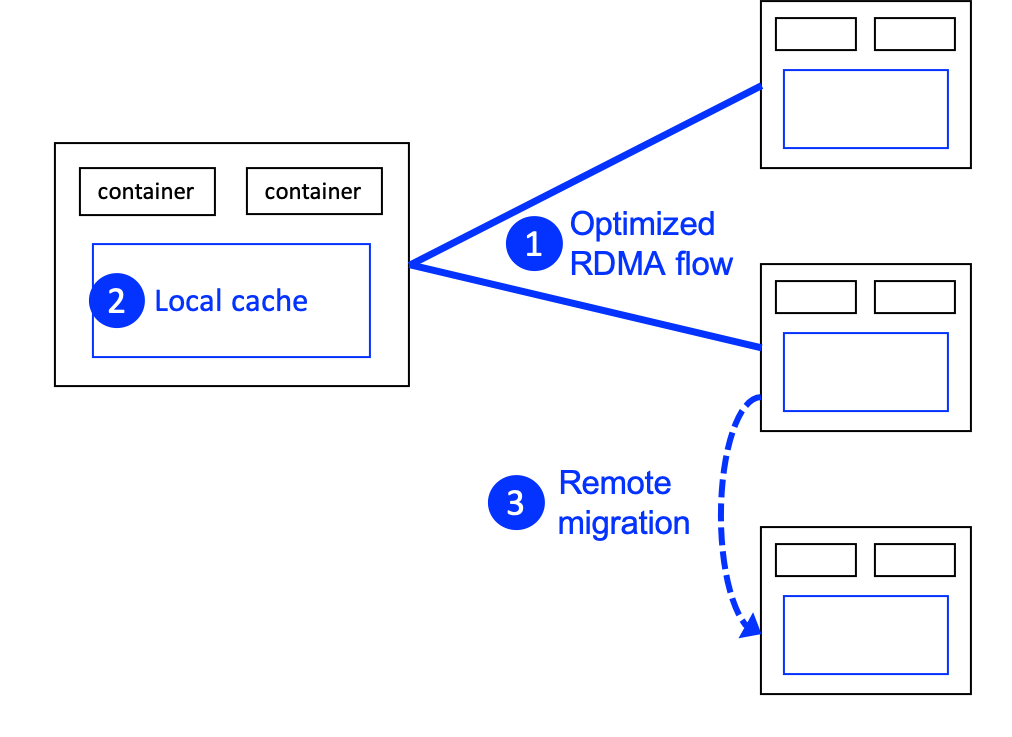}
\caption{Summary of contributions in Valet.}
\label{summary}
\end{figure}

We design and develop Valet to address the above challenges with three original contributions(Figure \ref{summary}). First, to reduce the hidden latency in the critical path, we redesign the data flow in the critical path by introducing a shared memory pool that works as a local cache to remote data. As a result, Valet shortens performance critical path and hides disk access scenarios unlike previous work~(\cref{latencyoverheadsolution}). Second, Valet utilizes idle node level (host) memory across containers via the node-coordinated shared memory pool. This helps to maximize local idle memory utilization and improves application performance on containers~(\cref{containerimbalancesolution}). Third, Valet provides an efficient remote memory reclaiming technique to minimize the impact of eviction from a remote node on the performance of local containers (local node). Valet achieves the remote memory reclamation by introducing a data migration protocol to move the least active block of data to a remote node of less memory contention. This also helps to maximize remote idle memory utilization across cluster~(\cref{evictionimpactsolution}).

We evaluated Valet with both memory intensive big-data workloads: Memcached\cite{Memcached}, Redis\cite{Redis}, VoltDB\cite{VoltDB} on YCSB\cite{YCSB}, and memory intensive machine learning workloads: GradientBoosting Classifier, Kmeans clustering, Random Forest Classifier, Logistic regression\cite{Scikit-learn}\cite{jia2014caffe}\cite{PowerGraph} and TextRank\cite{TextRank}. Using Valet, throughput improves by up to 282$\times$\ and latency decrease by up to 98\% over conventioal OS disk swap. Compared to existing representitive remote memory paging system such as nbdX\cite{nbdX} and Infiniswap\cite{Juncheng}, throughput improves up to 5.3$\times$\ and latency decrease by up to 80\% , demonstrating that Valet is an efficient memory orchestration framework for managing both idle host memory and idle remote memory, and maximizing peek time performance of memory intensive workloads in the presence of transient memory usage variations~\cite{xmempod}. 

In the rest of the paper, we first describe the problems of existing approaches and the challenges to be addressed in Section \cref{softwarechallenges}. We present an architectural overview of Valet in Section \cref{designoverview} and \cref{implementation}. We provide discussions in Section \cref{discussions} and experimental evaluations in Section \cref{evaluation}. Section \cref{relatedwork} presents the related work and section \cref{conclusion} concludes the paper.

\section{Software Challenges}
\label{softwarechallenges}
Before we go into discussions of software challenges in remote paging system, we try to define the term we use in this paper. In remote paging system, local node(or sender node) handles swapping traffic and remote peer node(or receiver node) allocates memory and registers RDMA Memory Region(MR) blocks as a memory donor for multiple sender nodes. Local node also has multiple peer nodes to distribute paging-out(or write) requests and to read data for paging-in(or read) requests.

\subsection{Latency Overhead in the critical path.}
\label{latecnyoverheadproblem}
In in-memory systems utilizing extremly fast DRAM and RDMA, design of critical path in I/O request accounts for the huge portion of overhead in the I/O performance. To understand the burden on latency, we build a prototype of network block device as a baseline. Typical design of RDMA based network block device uses one sided verbs to bypass the kernel at remote side. Before starting I/O operations, connection establishment and mapping to remote MRs are required. We choose dynamic connection and mapping mechanism. We apply power of two choices mechanism for dynamic connection and mapping node selection. Connection and mapping involve querying N remote nodes and selecting the most free node. It also needs address/route resolution, connection establishment and exchanging MR address and keys. Lastly, we add asynchronous disk backup on local side. These design choices are similar to the current state of art remote paging system\cite{Juncheng}. We measure latency of each operation to figure out the impact of the latency overhead in general cases. We set our block device as a partition and run FIO microbenchmark on it with the range of 128Kb block I/O size. Write size can be from 4KB up to 128KB and read size is 4KB for both disk and RDMA operations. We run over 10 thousand operations and take an average. Obviously, disk write has the biggest overhead as we expected but we also find out that the latencies for dynamic connection and mapping are not trivial as shown in Table \ref{latencycomparison}. 


\begin{table}[ht!]
\centering
\includegraphics[width=50mm]{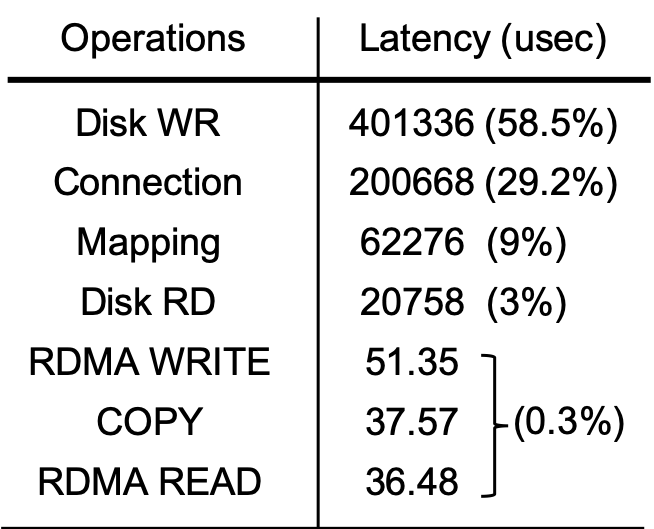}
\caption{Comparison of latency impact on the critical path in typical design of network block device. \textmd{Connection and mapping have significant overhead on the critical path.}}
\label{latencycomparison}
\end{table}

In existing design choices, we find that there are several contributory factors to the inefficiency. 
First, Performance critical path of I/O is tied with remote sending operation. In one-sided operation, I/O request ends when WC(Work Completion) is polled from CQ(Completion Queue). In two-sided operation, it ends with receiving the response message from receiver node. Second, another latency in the critical path is related to connection establishment and mapping. Connection might not be expensive because it happens only once per receiver node but mapping is. There are two approaches here. One is pre-mapping and the other is dynamic-mapping. Pre-mapping for all possible remote memory in peer nodes removes the mapping latency from the critical path but it is not scalable and also wastes too much resources for internal data structure and buffers that might not be used. Dynamic-mapping is scalable but mapping latency stays in the critical path. As shown in Table \ref{latencycomparison}, connection and mapping cost in the critical path are significant compared to RDMA operations and copy latency. Third, we observe disk access increases during connection and mapping setup because traffic has to be stored in somewhere while remote sending operation is blocked. Those data stored in disk will be accessed by read request later, which causes disk read activities.

\bigskip

\subsection{Container-wide Memory Imbalance}
\label{containerimbalanceproblem}

\begin{figure}[ht!]
\centering
\includegraphics[width=80mm]{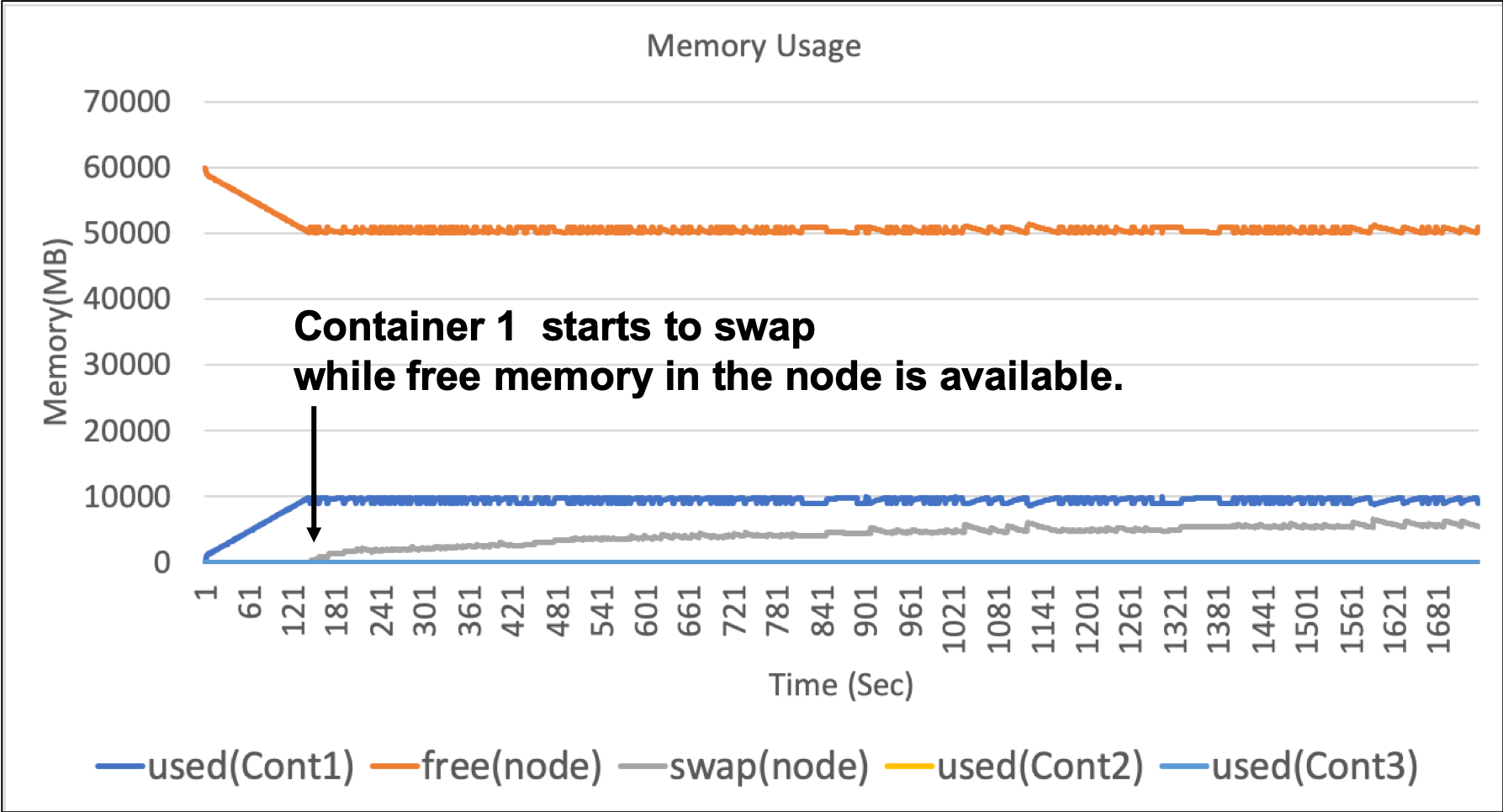}
\caption{Container-wide memory imbalance. \textmd{With the container memory limit setting, container cannot use more than its own limitation. We run 3 containers with memory limit in the node and measure memory usage while we run an application in container 1. Container 1 has 10GB memory limit. After 10GB is reached, container 1 suffers from swapping while unused memory remains in the node. Container 2 and 3 are idle at this moment.}}
\label{containerimbalance}
\end{figure}

\begin{figure}[ht!]
\centering
\includegraphics[width=70mm]{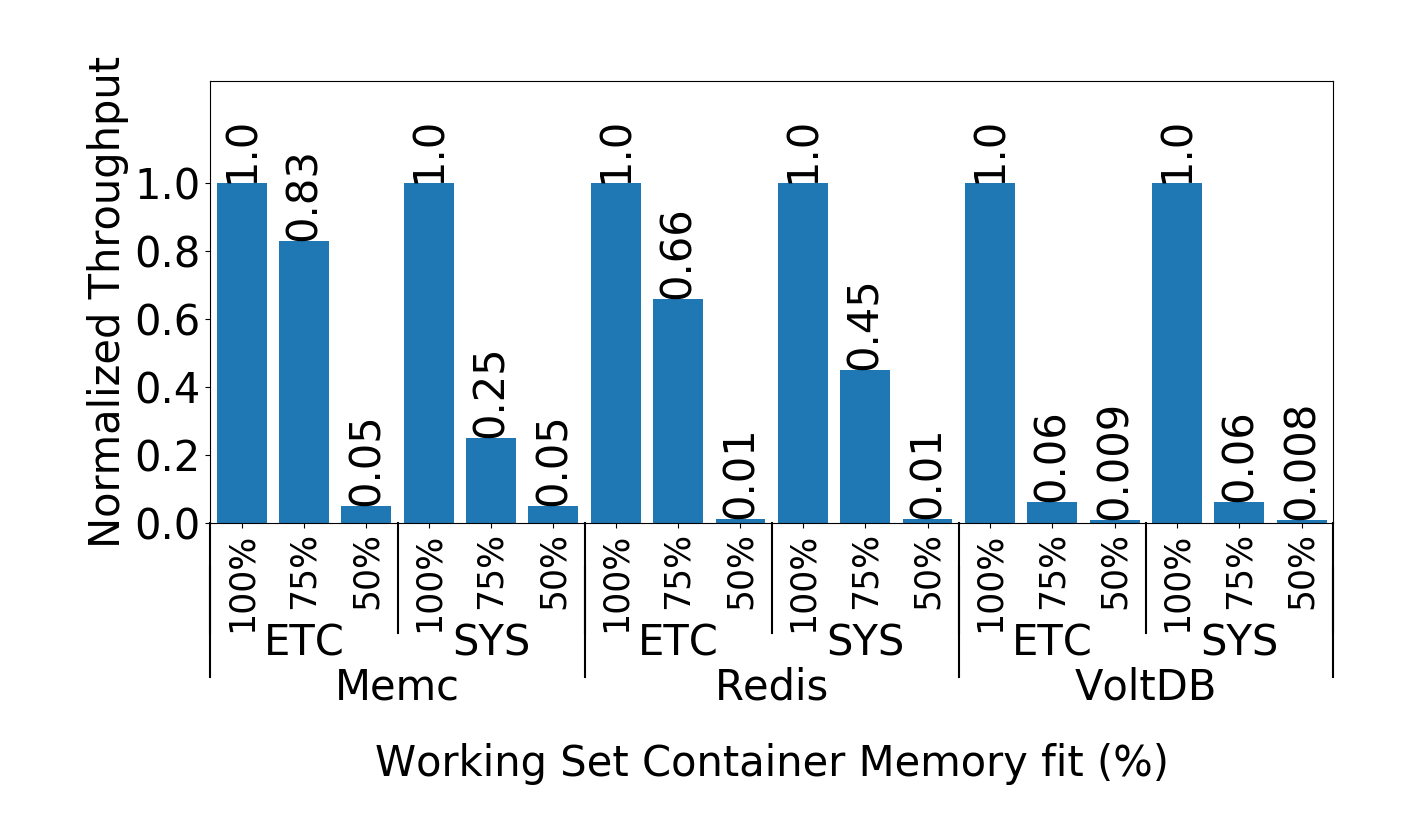}
\caption{ Applications performance with the setting in Figure \ref{containerimbalance}. \textmd{Applications suffer from performance degradation while unused memory remains in other containers.} }
\label{contimbalancemeasure}
\end{figure}

OS virtualization is a commonly used technology in many cloud servers and datacenters to provide isolated computing environment. There are two ways to set container's memory constraints. One is to set a limit of memory to each container. Applications on this container can use memory within the limitation. The other is to set unlimited. With unlimited settings, one container can consume all the memory in a node. Then, others running later suffer from performance degradation by swapping to disk. With memory limit, container-wide memory imbalance exists among multiple containers on the same node because cloud systems typically serve heterogeneous guest application workloads and it shows heterogeneous data access patterns during runtime\cite{Ling}. Figure \ref{containerimbalance} shows memory imbalance situation where container 1 suffers from swapping while free memory remains on the node. In Figure \ref{contimbalancemeasure}, We run Memcached, Redis, VoltDB with varying the memory limitation of the container. For workload, we use Facebook simulated workload\cite{Facebookworkload} ETC(95\%GET and 5\%SET) and SYS(75\%GET and 25\%SET) by using YCSB\cite{YCSB}. Performance severely decreases due to swapping while unused local memory remains in the node. Previous approaches\cite{nbdX, Juncheng} are not free from this container-wide memory imbalance problem.


\begin{figure}[!htb]
\begin{minipage}[t]{0.5\textwidth}
\centering
\includegraphics[width=0.7\linewidth]{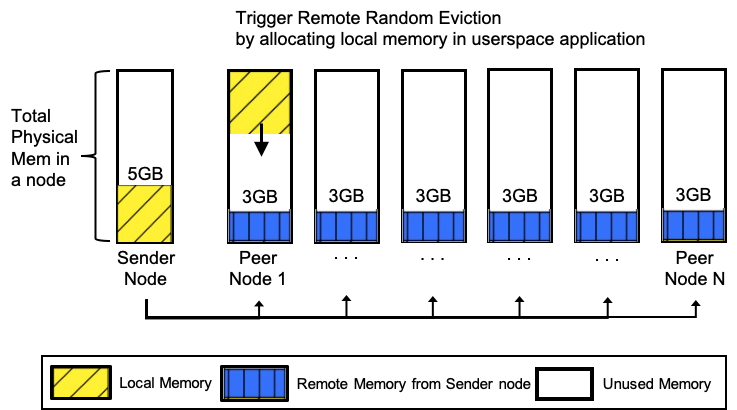}
\caption{Experiment setup. \textmd{To figure out remote eviction impact on sender node, We run 6 peer nodes for a sender node. Container in the sender node has 5GB limit. When 5GB limit is reached in the sender node, about 18GB workload is evenly distributed into 6 peers in the cluster. We run native applications on M peer(s) at each run to allocate all free memory and cause the remote eviction, where M is 1 to 6. Local memory denotes consumed memory on both sender and remote peer node and remote memory denotes data from sender node.}}
\label{evictionimpactsetup}
\end{minipage}

\begin {minipage}[t]{0.5\textwidth}
\centering
\includegraphics[width=0.7\linewidth]{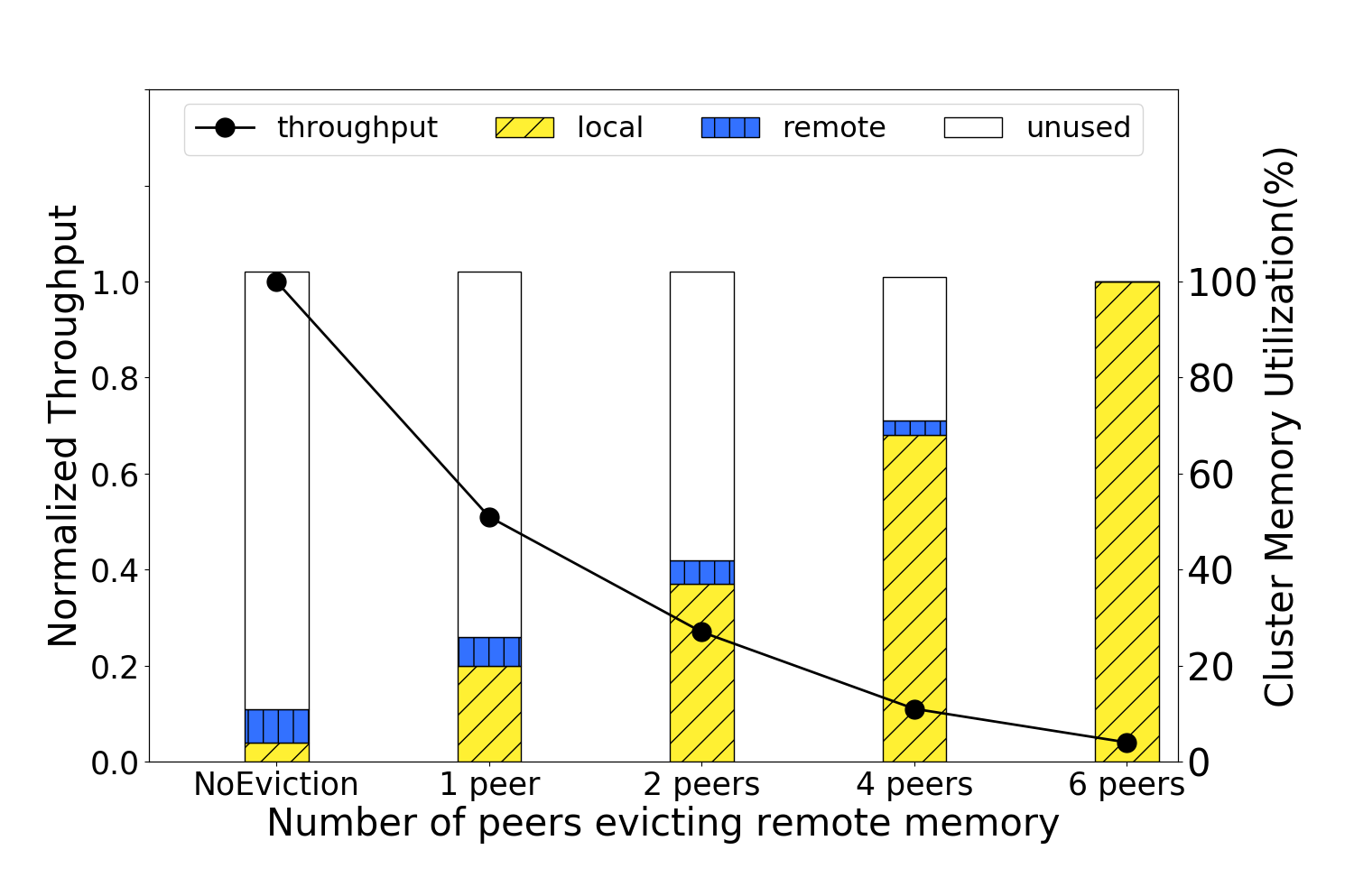}
\caption{Remote eviction impact and imbalanced cluster-wide memory utilization. \textmd{Line represents normalized throughput of Redis on sender node and bar represents cluster memory utilization of 6 peers. Remote eviction happens from 1 to 6 peers at each run(Figure \ref{evictionimpactsetup}). Evicted data from peer nodes causes significant performance degradation while unused memory in other peer nodes is not fully utilized(e.g. when only 1 peer evicts all remote memory, it shows 50\% decrease in throughput of Redis on sender node).}}
\label{evictionimpactexp}
\end{minipage}
\end{figure}

\subsection{Remote Eviction Impact}
\label{evictionimpactproblem}

Remote eviction happens due to shortage of free memory when applications in the remote node call for memory. When remote memory eviction happens, performance impact on sender node is inevitable because remote memory is simply deleted from the peer node. Later, all read requests to those deleted data are served from disk in the local node. If the deleted data is highly active one, the impact on sender node is even worse. Another problem is that, finding the most inactive victim is costly. Typical way of handling this is to query write/read activity to multiple sender nodes. This unnecessarily increases communication latency to query sender nodes if the remote memory block is inactive. If the number of queries gets bigger to find the victim well, communication latency increases linearly. In turn, it results in memory pressure on native applications on the peer node due to slow eviction process. Regarding scalability perspective, the impact on sender node due to eviction increases as workload increases. The more pages reside on the peer node, the higher risk of eviction exists and the impact is larger. We measure eviction impact with 23GB workload. We first run Redis with SYS workload to populate 6 peers(See Figure \ref{evictionimpactsetup}). Then, we run native application in the peers until it consumes all free memory. Then, receiver module that manages remote memory evicts remote memory by randomly selecting 1GB sized remote memory block at a time until all blocks are evicted. Figure \ref{evictionimpactexp} shows throughput of Redis and cluster-wide unused memory. It shows that eviction causes significant performance loss on the sender node and it becomes worse as the amount of evicted data increases. It also shows that idle memory in the cluster remains unutilized while throughput severely decreases. Addressing remote eviction impact is critical to achieve scalability in distributed in-memory systems.

\section{Design Overview}
\label{designoverview}

\subsection{Design considerations}
\label{designprinciple}

\hspace{\parindent}\textbf{Maximize CPU utilization}
Valet employs asynchronous I/O to maximize CPU utilization. Multi queue block I/O mechanism is working with multiple threads.

\textbf{Critical path optimization}
Valet achieves shorter latency by optimizing performance critical path. With host-coordinated local mempool, dynamic connection, mapping to remote RDMA MR and local disk access are hidden from the critical path.


\textbf{Utilize unused memory}
Valet utilizes unused memory both in local and remote memory. Valet tries to utilize the unused memory that is managed by host-coordinated memory pool in a local node first. It exploits container-wide memory imbalance and manages free memory that is not used by other containers. This maximizes idle memory usage in a local node. Valet also utilize unused remote memory in remote nodes by dynamically registering RDMA MR(Memory Region). Local node spreads paging-out data to multiple remote node based on the amount of free memory. 

\textbf{Reclaim memory efficiently}
Reclaiming memory is also crucial for native applications running on both local and remote nodes. Host-coordinated local mempool dynamically expands and shrinks according to the amount of free memory in the local node. Remote RDMA MR also expands and shrinks according to the free memory on the remote node. Valet also provides migration protocol for remote eviction. It migrates victim data block to other less-memory-pressured nodes. This also maximizes idle memory usage in remote nodes. 

\textbf{Reliability}
Valet uses staging queue and reclaimable queue to maintain the data consistency between local and remote nodes. Unlike parallel reading(paging-in), writing(paging-out) is serialized for data consistency. Valet also provides replication across remote nodes for diskless design. We prefer replication over disk backup. Even though SSD is faster than rotational disk, RDMA is still more than 20 times faster than SSD\cite{Orion}.

\textbf{Scalability}
Scalability is essential for Valet to process large amount of workload. Valet scales well with multiple remote nodes and distribute workload across remote nodes. Valet also keeps low latency while workload increases. Valet acheives this by removing bottlenecks in the data path.

\begin{figure}[ht!]
\centering
\includegraphics[width=90mm]{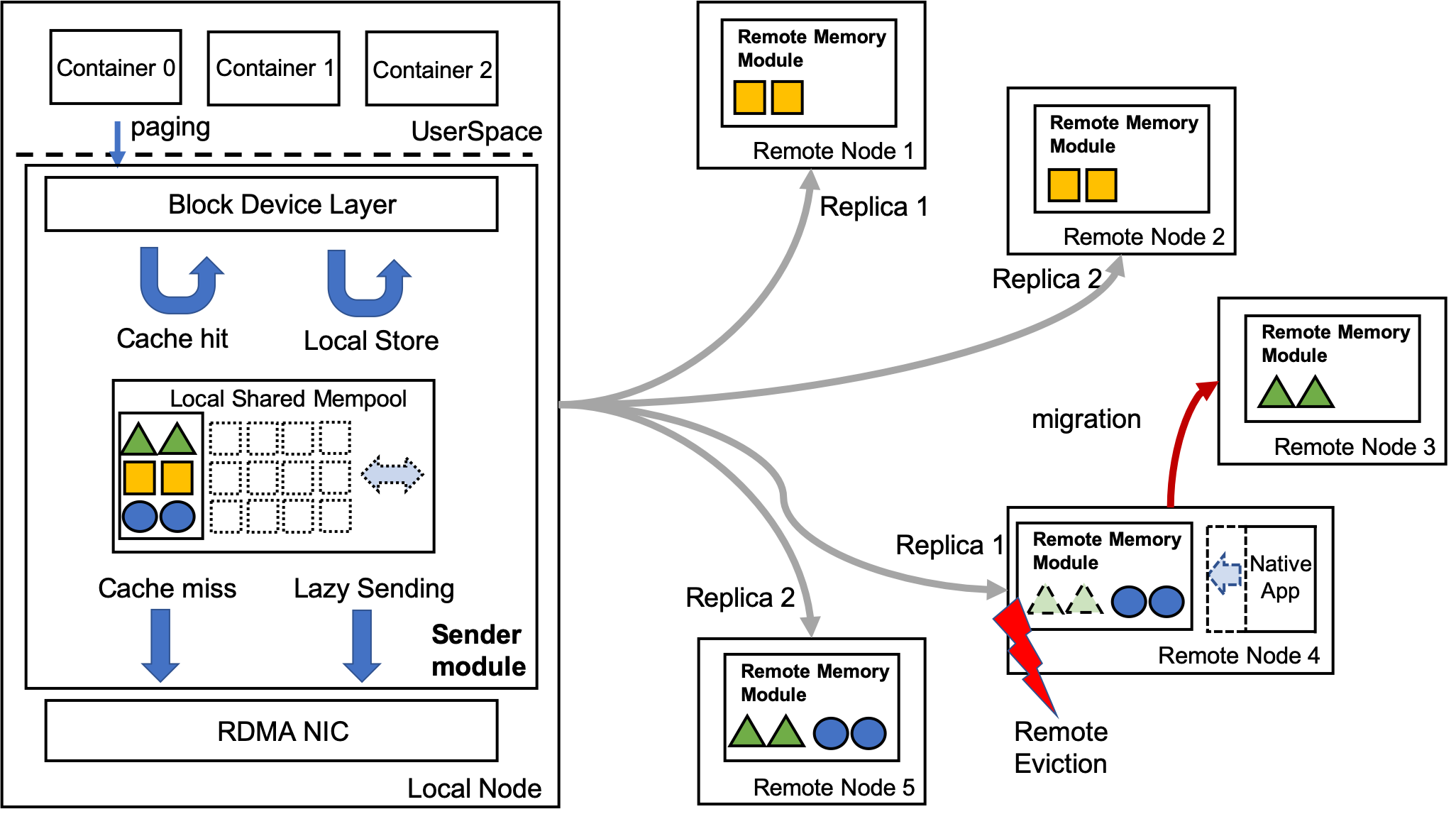}
\caption{Overall software organization of Valet.}
\label{overall}
\end{figure}

\subsection{Software organization}
In Figure \ref{overall}, we show overall software organization in Valet. Valet uses symmetric model. Each node can be a sender and a memory donor(receiver) at the same time although it is not a requirement. Sender module takes swap traffic. Receiver module(Remote Memory module) manages MR blocks as remote memory. Sender node can allocate remote memory across multiple remote nodes. Remote node can serve multiple sender nodes in the cluster. 

Valet ends a write request after storing pages in a local shared memory pool(local mempool in short for the rest of the paper). Pages that are stored in the local mempool will be sent out to remote nodes later asynchronously. For read requests, Valet tries to find the page from local mempool first and reads from a remote node if cache is missed. The local mempool extends and shrinks to maximize local idle memory utilization. 

Valet tries to spread data evenly across the cluster. If remote eviction happens in a remote node it moves remote memory block to less-memory-pressured node. This maximizes cluster idle memory utilization. Detailed discussion of components in Valet can be found in section \ref{implementation}.

\subsection{Performance critical path optimization}
\label{latencyoverheadsolution}

\begin{figure}[ht!]
\centering
\includegraphics[width=90mm]{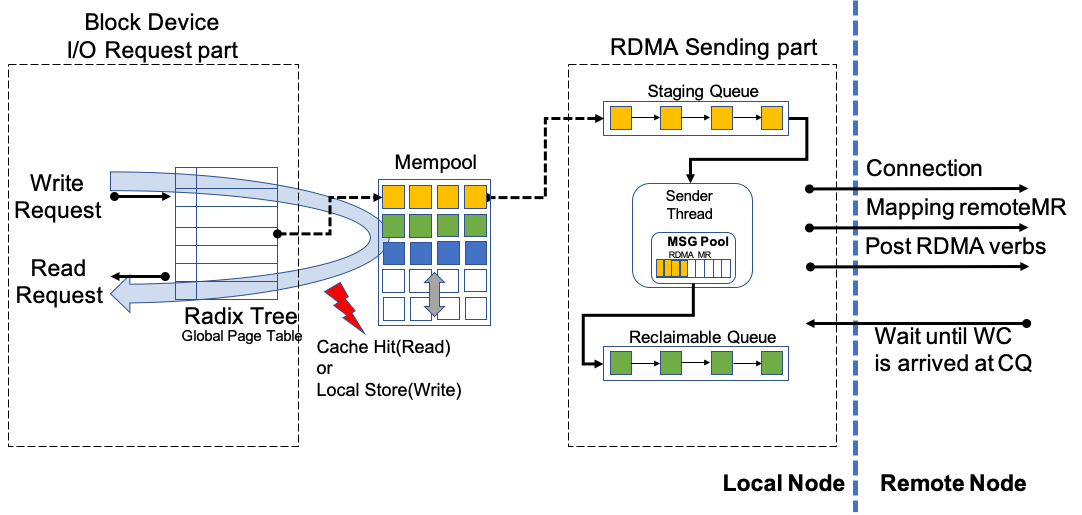}
\caption{Redesign critical path. \textmd{With performance path optimization, RDMA Sending part is detached from the performance critical path. Connection, mapping to remote RDMA MR and RDMA verb operations are hiding from performance critical path. For read, Valet shortens read critical path when local cache hit is made. }}
\label{separate}
\end{figure}

\noindent
\textbf{Redesign Critical Path.}
Valet redesigns performance critical path by having host-coordinated local mempool. For write case, as soon as it stores pages into local mempool, it can immediately end the I/O request and accomplish shorter write latency. The rest of the remote sending operations are done after the data is written to the local mempool and mempool starts servicing for read request(Figure \ref{separate}). 

\begin{figure}[!htb]
\begin {minipage}[t]{0.23\textwidth}
\centering
\includegraphics[width=1\linewidth]{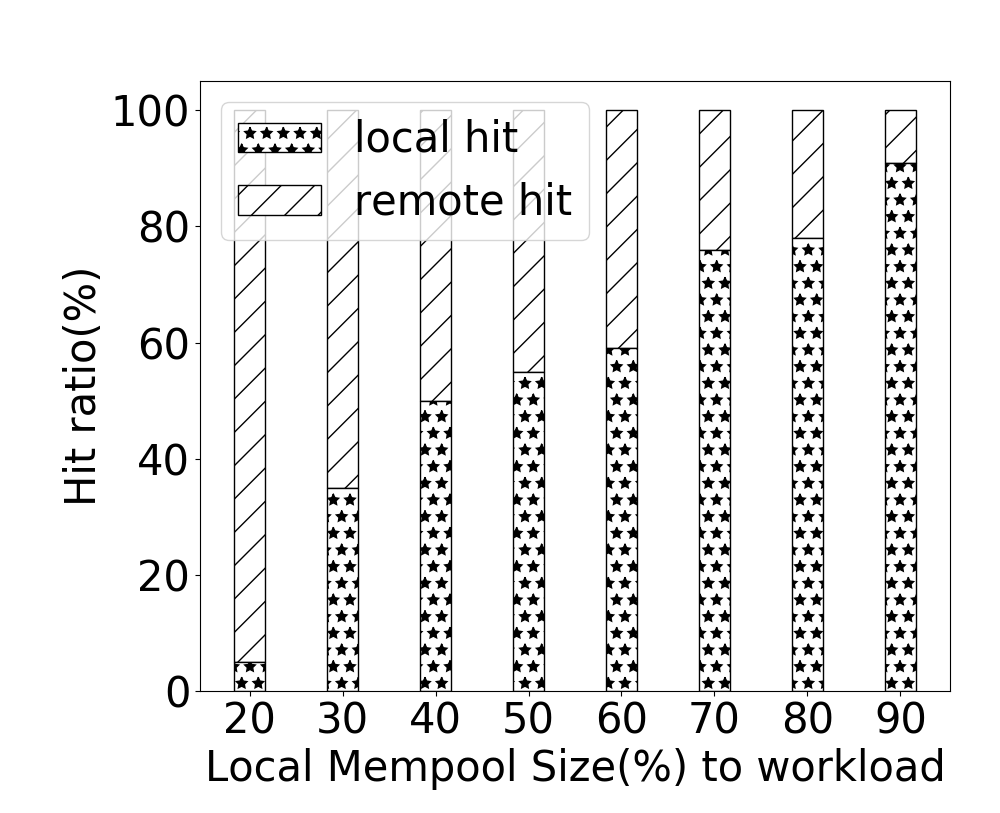}
\caption{Local and Remote hit ratio comparison with various local mempool size. \textmd{Local hit ratio increases as local mempool size increases.}}
\label{localmempoolsize}
\end{minipage}\hfill
\begin {minipage}[t]{0.23\textwidth}
\centering
\includegraphics[width=1\linewidth]{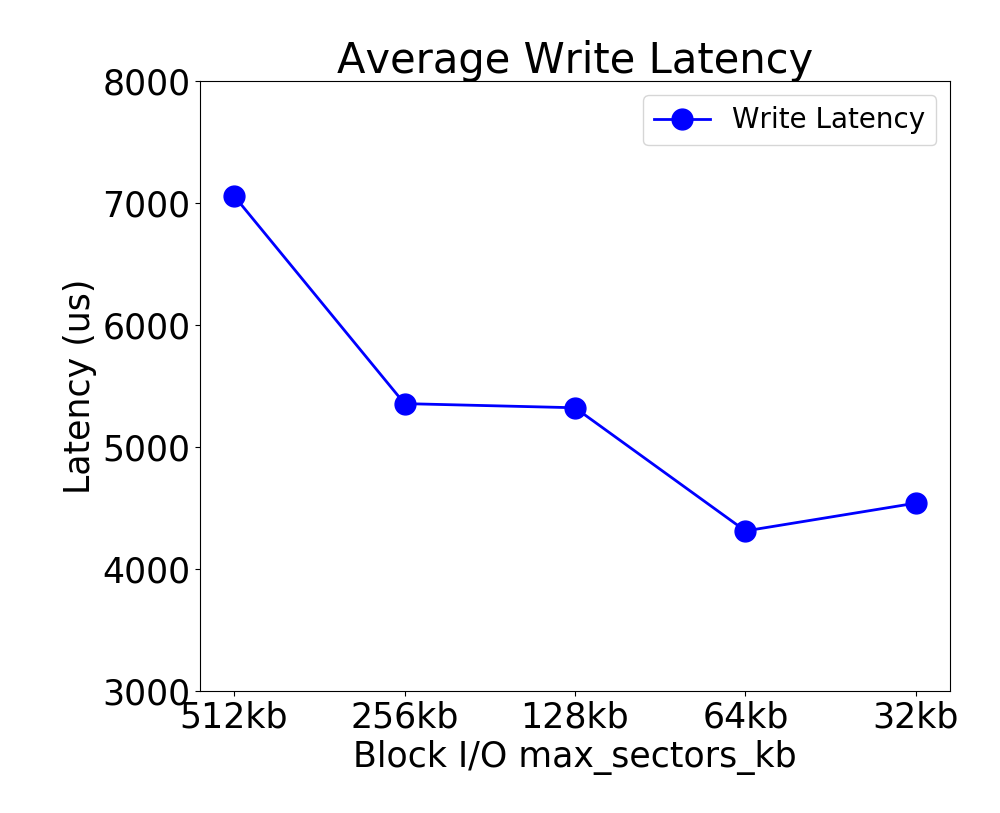}
\caption{\textmd{With performance path optimization, application write latency decreases as the Block I/O size decreases because only I/O request part remains in the critical path.}}
\label{maxsectors}
\end{minipage}
\end{figure}

Local mempool also functions as a cache for remote data. If data resides in local mempool(cache hit), remote access is not needed. Performance benefit(\cref{containerimbalancesolution}) gets larger when local mempool size increases as local hit ratio increases(Figure \ref{localmempoolsize}).



\bigskip
\noindent
\textbf{Pipelining the local mempool in the critical path.}
Valet also hides connection and mapping of remote MR(Memory Region) latency from the write critical path. This design helps to remove cases that make read latency high too. During connection to a remote node and mapping to a remote memory block, I/O request traffic should be redirected. Valet stores I/O traffic in the local mempool instead of disk. By directly serving read request from the local mempool, it can avoid long read latency due to disk access that is caused by delay of connection and mapping. After connection and mapping are done, local-stored data is sent to remote node to reduce the memory pressure on local mempool.


\bigskip
\noindent
\textbf{Flexible design for input I/O and RDMA buffer size}
Unlike previous designs, Valet's I/O request size is not tied with RDMA MR size. Previous design approaches share the same buffer for RDMA MR and disk writing to avoid extra copies. It is also bounded by max size of hardware disk I/O capacity. \textbf{max\_sectors\_kb} determines the number of pages in one Block I/O request. If system has M kb \textbf{max\_hw\_sectors\_kb} of hard disk, the size of Block I/O and RDMA MR size for remote paging system are bounded by this hard disk's physical limitation. Valet can set different value for Block I/O and RDMA MR size regardless the hard disk's block I/O size limitation if one wishes to add disk backup. The benefit of having different size between Block I/O and RDMA MR is of having a chance to optimize according to various desires. Generally speaking, block I/O size affects the write latency because it adds latency in the critical path while copying pages from Block I/O buffer to RDMA MR. If Block I/O size is set by large number, a Block I/O request has more pages and, in turn, it takes longer time to copy. If the size is small, it takes less time to copy, which leads to shorter latency. See Figure \ref{maxsectors}. write latency decreases as block I/O size gets smaller. The latency of 32KB is slightly higher than 64KB because of CPU burden due to too many small requests. If RDMA MR size is small, the number of RDMA I/O should increase to send the same amount of data. it may cause WQE(Work Queue Entry) cache miss due to many WQEs injecting to RDMA NIC. It is discovered in previous research\cite{Dragojevic} that many WQEs cause WQE cache misses in NIC. Valet takes the advantage at this point. Valet can set small size of block I/O to get low latency and use message coalescing and batch sending with large size of RDMA MR to avoid WQE cache miss. 


\subsection{Utilizing unused memory}
\label{containerimbalancesolution}

\textbf{Container-wide memory imbalance and Lazy Sending}
Local mempool provides a chance to use idle local memory that is not used by other containers by combining them into the local mempool. The local mempool shrinks when the amount of free memory goes below the user defined threshold to guarantee the certain amount of free memory in the node. Then, local pages in the mempool are sent to remote nodes and reclaimed. Before this page replacement happens, this lazy sending scheme best tries to utilize unused local memory and lower the memory pressure on the remote node. Local mempool can grow again when the free memory in the local node goes above the threshold for expansion.

\begin{figure}[!htb]
\begin{minipage}{0.23\textwidth}
\centering
\includegraphics[width=1.1\linewidth]{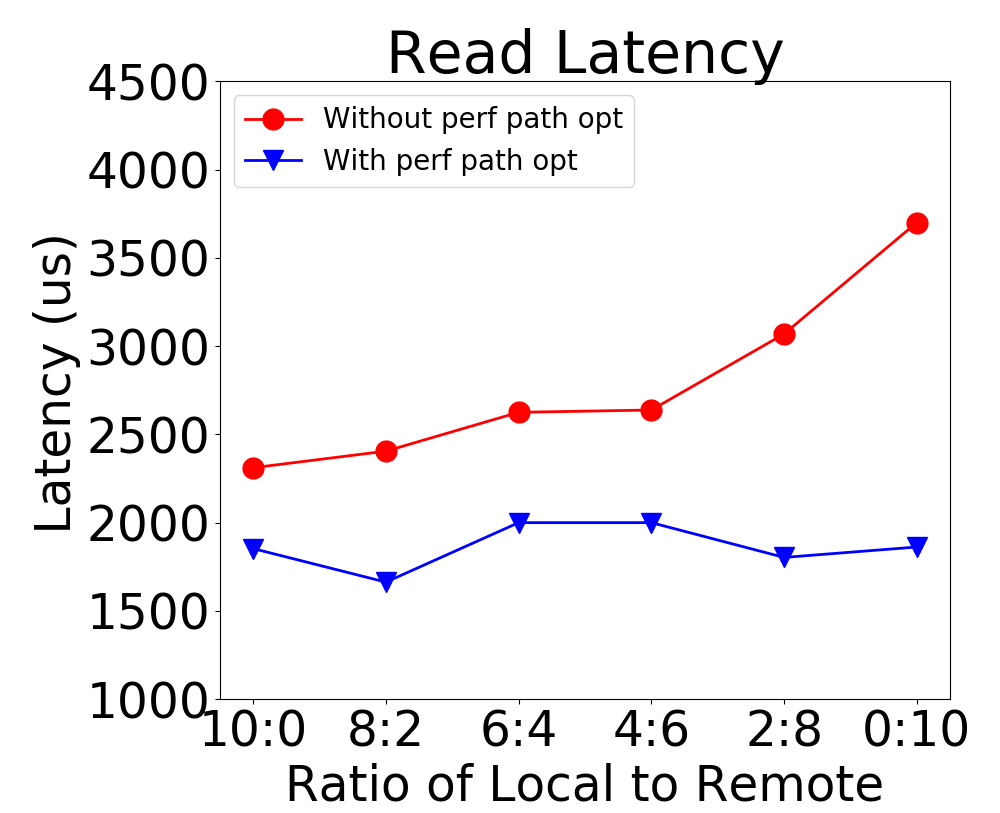}
\end{minipage}\hfill
\begin {minipage}{0.23\textwidth}
\centering
\includegraphics[width=1.1\linewidth]{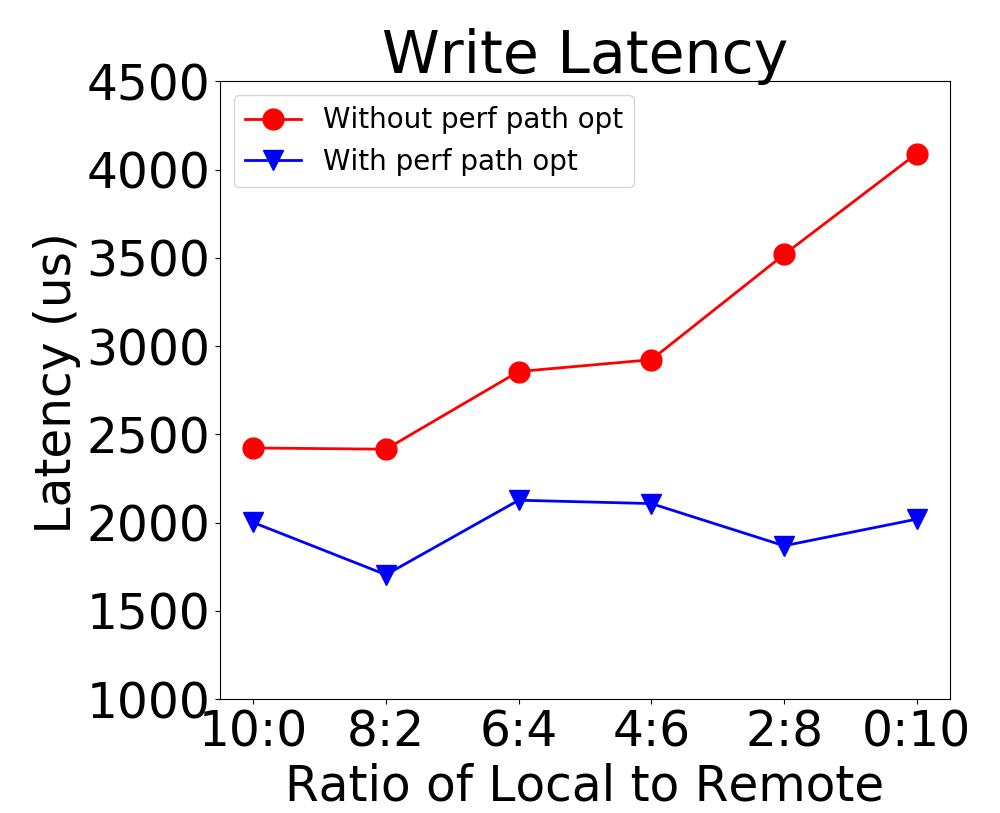}
\end{minipage}
\caption{Latency comparison with and without critical path optimization. \textmd{With performance critical path optimization, application latency stays stable regardless of the various ratio of local to remote memory.}}
\label{localvsremote}
\end{figure}

\bigskip
\noindent
\textbf{Impact of the size of mempool}
Since the local mempool can dynamically expand and reduce adaptive to the workload dynamics, we first measure the percentage of local hit over remote hit with various size of local mempool to figure out the local mempool's contribution to local hit. As shown in Figure \ref{localmempoolsize}, large size of mempool gets more local hit. If local mempool size decreases, it gets more remote hit. Application latency stays stable with mempool compared to the one without critical path optimization. In Figure \ref{localvsremote}, we run VoltDB SYS workload with 10 million records and 10 million queries under various ratio of local memory to remote memory by setting container memory limit. 10:0 denotes I/O is served only in local memory and 0:10 denotes only in remote memory. 


\subsection{Reclaiming remote memory}
\label{evictionimpactsolution}

\begin{figure}[ht!]
\centering
\includegraphics[width=80mm]{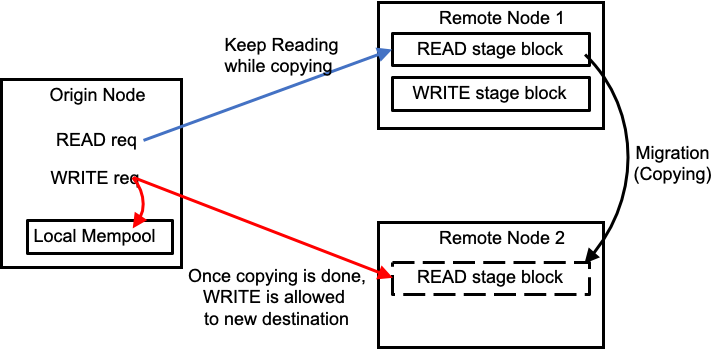}
\caption{Read requests are allowed to access remote MR block while copying but write requests stay in the local mempool. \textmd{By choosing the least-active MR block as a victim(likely idle or read stage), sender node can lower the memory pressure on the local mempool due to few writes.}}
\label{whyoldchunk}
\end{figure}

\textbf{Data migration instead of delete}
Valet uses migration protocol when remote data eviction happens in remote node. The major benefit of migration is that it does not hurt the throughput of sender node that maps the data. In order to avoid the I/O blocking during the migration, we allow read requests while migration is in progress. Regarding data consistency concerns between source and destination due to write requests during migration, a local mempool in the sender node can hold the write requests in the local mempool. All the new write requests to the migrating data stay in the staging queue until migration is done. Since these queued write requests are stored in local mempool, read requests to the data are guaranteed to read the latest data by reading from the local mempool. Once migration is done, the sender node can write/read to/from the new destination. Write requests in the staging queue can also be sent out to the new destination(Figure \ref{whyoldchunk}). Detailed discussion about consistency is in section \cref{dataconsistencysection}.

\begin{figure}[ht!]
\centering
\includegraphics[width=80mm]{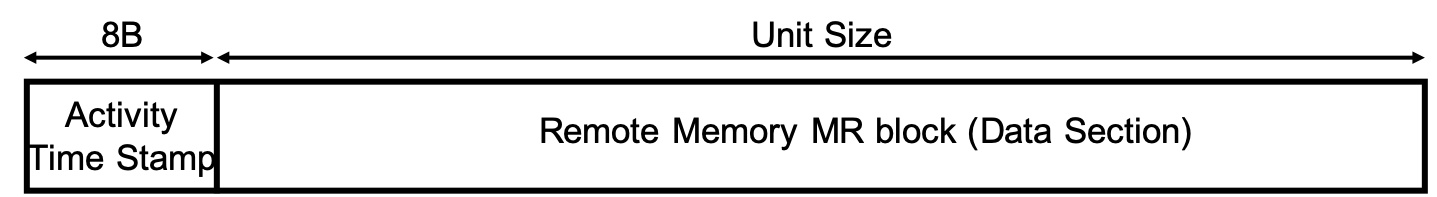}
\caption{Format of MR block on remote node. \textmd{Tag information is included to calculate Non-Activity-Duration at eviction.}}
\label{tag}
\end{figure}

\begin{figure}[ht!]
\centering
\includegraphics[width=80mm]{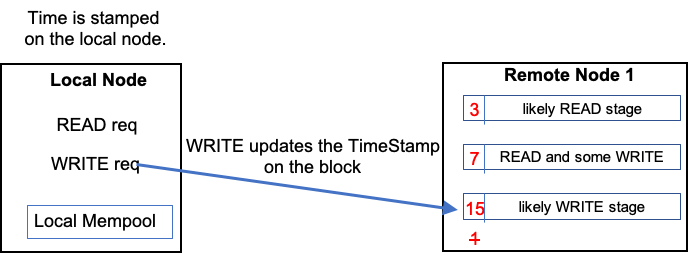}
\caption{TimeStamp on the MR block is updated by write request. \textmd{Then, this block becomes the most-active one in the node. In this figure, the number denotes conceptual last write activity timestamp for each block. The block that has 15 becomes likely the most active block due to recent update. Compared to others, 3 is the most likely read stage due to the longest Non-Activity-Duration among three}}
\label{chunkupdate}
\end{figure}

\bigskip
\noindent
\textbf{Activity-based Victim Selection on remote node}
Unlike read performance, write performance during migration relies on the capacity of local mempool because local mempool is responsible for holding writing requests to the migrating data on MR block during migration. Finding the least-active-MR-block as a victim is crucial factor to lower the memory pressure on the local mempool. To find the least-active-victim, we propose an activity-based victim selection algorithm. We calculate duration since last update for each MR block on the remote node. 
\begin{align*}
\textbf{Non-Activity-Duration} = Time_{\mathrm{cur}} - Time_{\mathrm{last\_activity}}
\end{align*}
Every MR block on the remote node has small metadata tag and the last write activity is timestamped(See Figure \ref{tag}). This last active timestamp is updated when this MR block is updated with write requests from its sender node(See Figure \ref{chunkupdate}). Non-Activity-Duration for each MR block will be calculated at the time of eviction process.

Through our observation of write pattern from various workload, we find that the activity cycle of the remote memory block starts with the heavy writes and becomes heavy read state and idle state as time goes by. If a remote MR block starts to receive write requests, it is highly likely followed by read requests. Once heavy read stage is passed, it becomes idle state. This activity cycle is likely repeating by updating with the write operation. The benefits of choosing the least-active-MR-block are of having low write-request-pressure on the local node while local mempool holds them during migration and reducing communication to query write activity to the sender node. The least-active-MR-block is highly likely to be idle stage. Valet can select this idle block by simply choosing the least-active one without querying to N sender nodes. Then, memory pressure due to holding write operations on the local memory is also limited.

\begin{figure}[ht!]
\centering
\includegraphics[width=90mm]{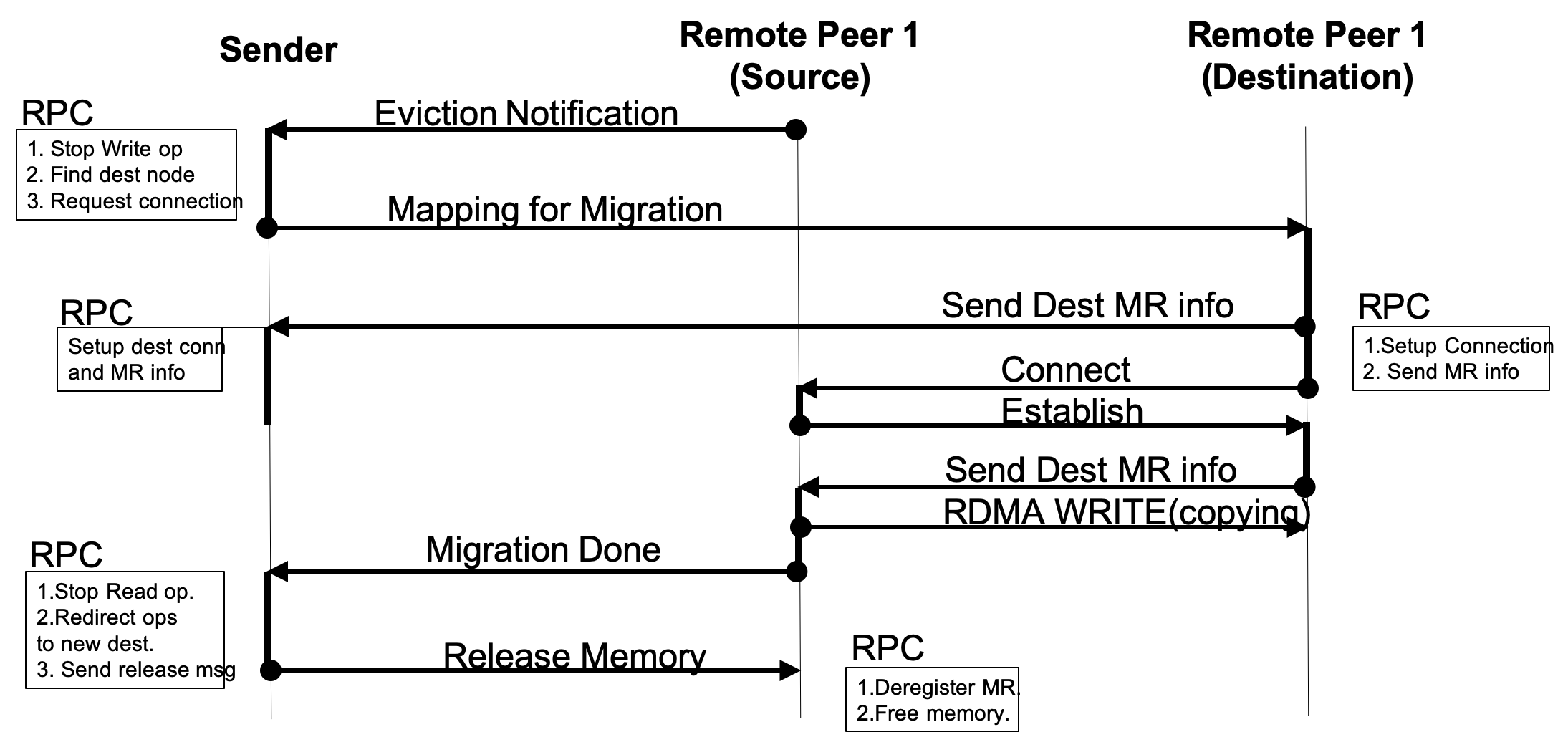}
\caption{Sender driven migration protocol}
\label{clientdriven}
\end{figure}

\bigskip
\noindent
\textbf{Sender driven migration protocol}
Migration protocol involves many message ping pong and remote procedures. In sender side, it should stop write requests before migration starts and prepare necessary setup with new destination information. In receiver side, source and destination nodes need to communicate each other and share necessary information for source and destination block including connection setup. We propose sender driven protocol(Figure \ref{clientdriven}). In sender driven protocol, sender node takes responsibility for control of the migration procedure and selects proper migration destination node. Receiver nodes are passive participants. Remote procedure is executed when it receives control message. This serialization leads to simple message control model. Extra control for message ordering is not required. Sender driven approach also gets benefit from pre-connection to counter parts. To determine a migration destination, sender node needs to query N candidate-remote peer nodes. If no connection is setup before, connection latency is directly added to critical path in migration procedure. However, if the number of mapped remote memory block is larger than the number of peer nodes, all connections are likely setup before the time of eviction because sender node evenly spreads workload to peers. This behavior makes all candidate-peer nodes to be connected in advance. 

\section{Implementation}
\label{implementation}

\begin{figure}[ht!]
\centering
\includegraphics[width=90mm]{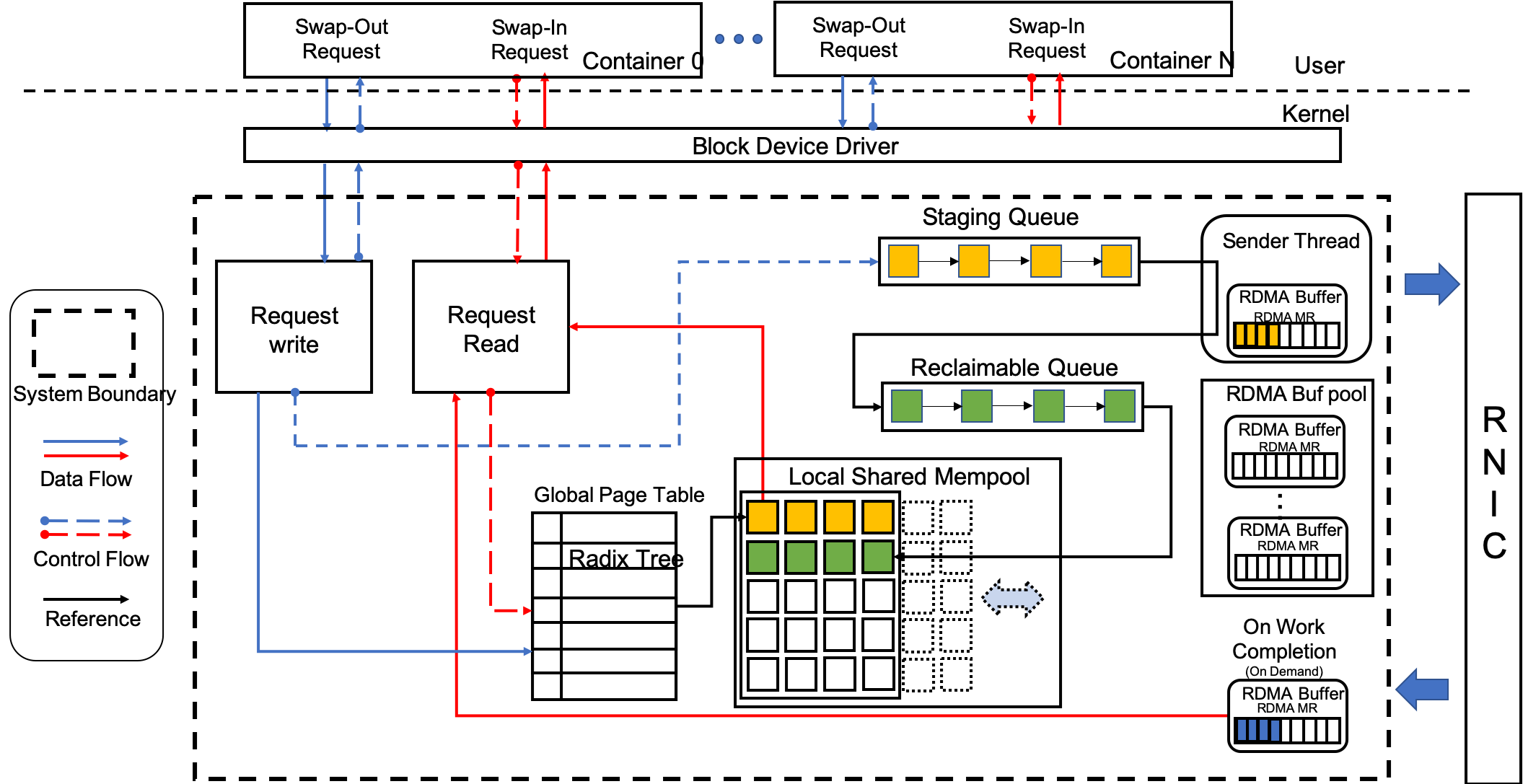}
\caption{Sender module architecture}
\label{clientmodule}
\end{figure}

\begin{figure}[ht!]
\centering
\includegraphics[width=85mm]{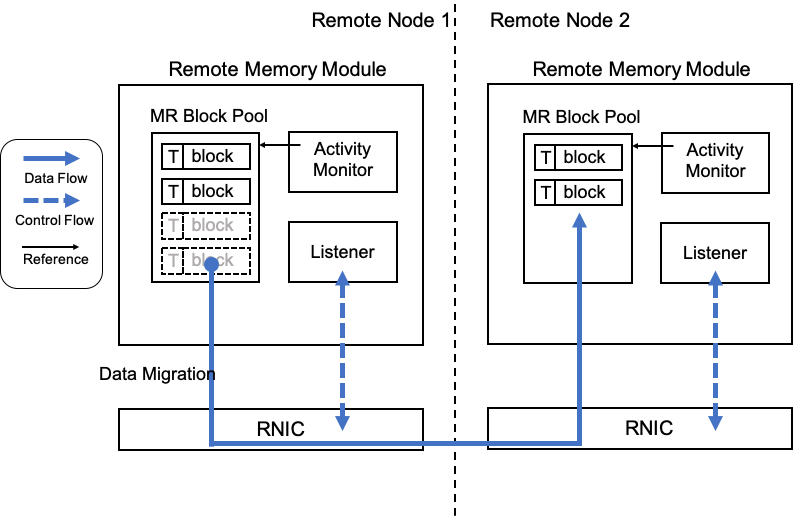}
\caption{Remote Memory module architecture. \textmd{Receiver module manages MR Block Pool. Activity monitor detects shortage of free memory and reports to sender node to initiate migration protocol. Then, source and destination receiver modules carry on the migration protocol}}
\label{servermodule}
\end{figure}

\subsection{Sender Module}
\label{sendermodule}

\textbf{Global Page Table}
Main role of GPT is to map the offset of the page to the reference of the pages in local mempool. Radix Tree is used to implement GPT. Radix Tree is wide and shallow structure tree. It is as fast as accessing to 1-dimensional array, which is the simplest design that GPT can be. Unlike array-based GPT, RadixTree-based GPT does not need to allocate the whole structure in advance. It can grow and shrink dynamically. This aspect more fits to our desire for scalable design. We use simple rule to locate a page. If a page reference exists in the GPT, it points to the local page. Otherwise, it indicates that the page does not exist in local memory. It then needs to read from remote memory by posting a READ verb. This simple design helps to avoid a lock contention on GPT update by removing the need for marking page existence on the GPT.

\bigskip
\noindent
\textbf{Dynamic Local Memory Pool}
Our mempool design is different from Linux Mempool implementation in several ways. Linux Mempool always tries to allocate memory first even if it has unused pre-allocated memory in the mempool. Pre-allocated pages in the Linux Mempool are only used when allocation is failed. It doesn't give a benefit of pre-allocation but gives a guarantee of allocation. In our design, we pursue three main rules. First, we want to avoid memory allocation burden on the critical path. Second, we want to have guaranteed amount of memory but use them first to minimize memory allocation latency in the critical path. Third, we want to have a flexible size of mempool based on the availability of free memory in the system. Figure \ref{mempool} shows the difference between them. Valet utilizes pages in pre-allocated mempool first and it can be extended or shrunk. The minimum size of the mempool is decided by user defined value \textbf{min\_pool\_pages}. With no user definition, if usage of mempool reaches 80\% of the current mempool size, Size grows on demand. It stops growing when it reaches to either \textbf{max\_pool\_pages} threshold or 50\% of the total free memory on the host node. Whichever smaller will be taken. If containers allocate memory and the size of free memory on the host node shrinks, the local mempool also shrinks accordingly and stops shrinking when it reaches to \textbf{min\_pool\_pages}. \textbf{min\_pool\_pages} guarantees the minimum size of local mempool.

\begin{table}[ht!]
\centering
\includegraphics[width=70mm]{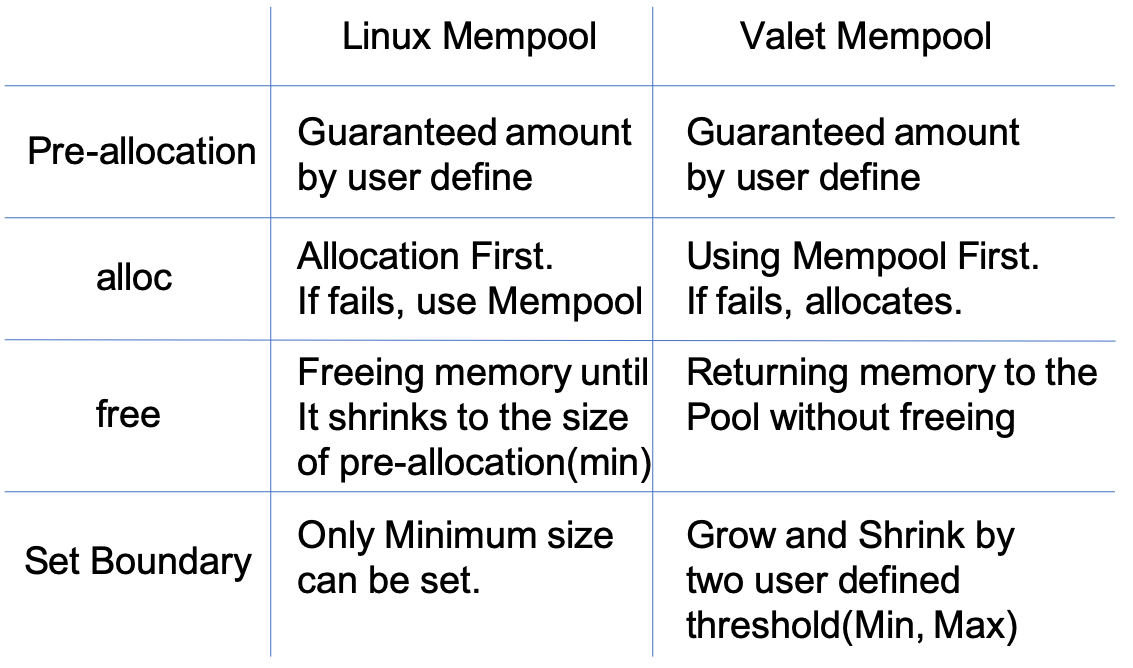}
\caption{Comparison between Linux Mempool and Valet Mempool implementation }
\label{mempool}
\end{table}

\bigskip
\noindent
\textbf{Local Mempool Page Reclaim}
Valet uses 24-byte sized \textbf{tree\_entry} structure to store page references and offset information from one Block I/O request, which represent one transaction in Valet. Staging queue and Reclaimable queue are responsible for tracking these entries that are already sent to remote and that aren't yet. When a write request arrives, the entry for the request is put into Staging queue. Remote Sender Thread takes an entry object from the Staging queue and sends pages to remote nodes. When message coalescing and batch sending are done, those page references are put into the Reclaimable queue. At this moment, pages tracked by Reclaimable queue are safe to be reclaimed because sending is done and a replicated copy is on the remote node. When local Mempool reaches 80\% of its size, mempool grows. If mempool cannot grow anymore, it starts to reclaim and provide free pages to new requests directly. For replacement policy, we use LRU in our prototype. Since reclaiming is just moving a page pointer, it takes only a few CPU cycles. 



\subsection{Remote Memory Module}
To reduce the CPU overhead on the remote peer node, Valet uses well-known one-sided RDMA verb to bypass kernel on the remote side. Remote Memory module maintains only necessary components and works as passive participant. The main purpose of Remote Memory module is to provide unit sized remote memory registered as MR to multiple sender nodes. Remote Memory module runs in user space and monitors free memory capacity in the remote node. Kernel space MR can utilize physically contiguous memory and reduce PTE cache miss in RNIC but allocation of large physically contiguous memory is challenging. User space MR requires RNIC to cache PTE to access the page because it uses virtually contiguous memory. However, user space allocation is much easier than allocation of large physically contiguous memory in kernel space. We use large MR blocks to reduce the number of MR mapping. Therefore, we choose user space receiver module design for MR block provider. It can dynamically expands and shrinks MR blocks based on the free memory. Remote Memory module also has listener to communicate with other receiver modules when they receive migration protocol messages. 


\subsection{How to track remote pages}
Valet provides block device interface. It can be registered as swap space or mounted as a partition with a linear address space. To track the location of remote pages, Valet defines global page address starting from 0 to the end of the user defined space size. This doesn't have to fit the remote memory capacity in the cluster. Then, this virtual address space spans across in the cluster. Mapping partitioned address space to remote peers happens on demand with round-robin or power of two choices. We use power of two choices in our prototype. Each unit sized address space and the same size remote memory blocks on remote nodes is dynamically mapped and internal data structure tracks this mapping information.

\section{Discussions}
\label{discussions}

\subsection{Fault Tolerance}

\textbf{Remote node failure.}
Valet provides several options for fault tolerance. Either remote replication and local disk backup or mixed approach can be selected as one wishes their fault tolerant level(Table \ref{fault_tolerance}). Each combination provides different semantic when remote node failure occurs.

\bigskip
\noindent
\textbf{Local host node failure.}
For permanent data store in local host, disk backup option is provided. Then local host writes backup on disk either always or only when writing to remote fails. In paging system example, we provide the same semantic to other paging systems when local node fails.

\begin{table}[ht!]
\centering
\includegraphics[width=70mm]{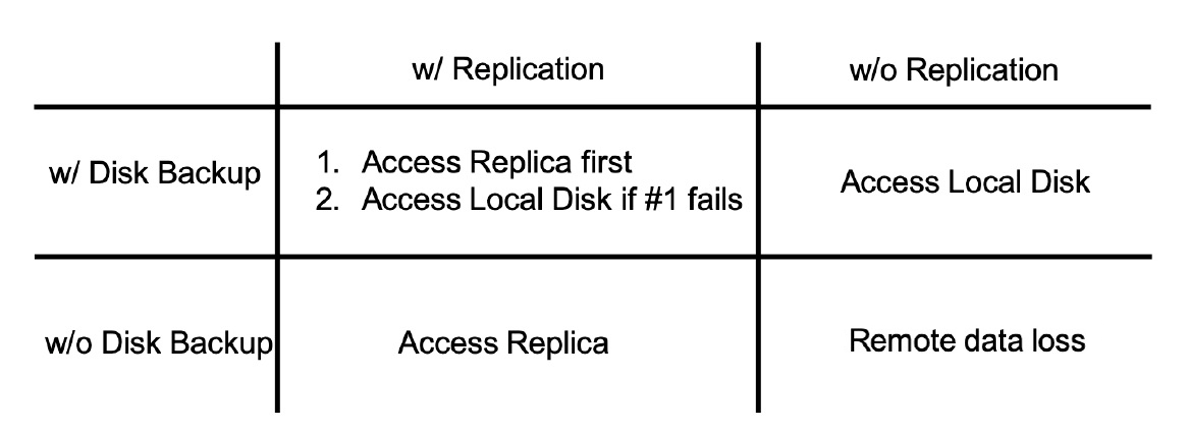}
\caption{Different level of fault tolerance is provided by combination of replication and disk backup}
\label{fault_tolerance}
\end{table}

\subsection{Data Consistency}
\label{dataconsistencysection}

\textbf{Between local memory and remote replicas.}
An incoming write request’s write sets are enqueued into Staging queue as the data is written to the local memory. If an incoming read request finds a page in the local mempool, it is always served from the local mempool directly. The remote pages are accessed only when local mempool does not have the pages due to reclaiming. This guarantees incoming read requests always get the most updated data. Remote Sender Thread takes write sets from Staging queue and sends out to remote nodes in incoming order. Once WC is received, bitmap for the remote page indicates that remote page is ready to read. This guarantees remote node has the same data when it is read. When remote sending is done, the write set is removed from the Staging queue and enqueued into Reclaimable queue. Page slots in the local mempool are reclaimed only through Reclaimable queue. Then only page slots that has replica on the remote node are guaranteed to be reclaimed for the next use.

\begin{figure}[ht!]
\centering
\includegraphics[width=90mm]{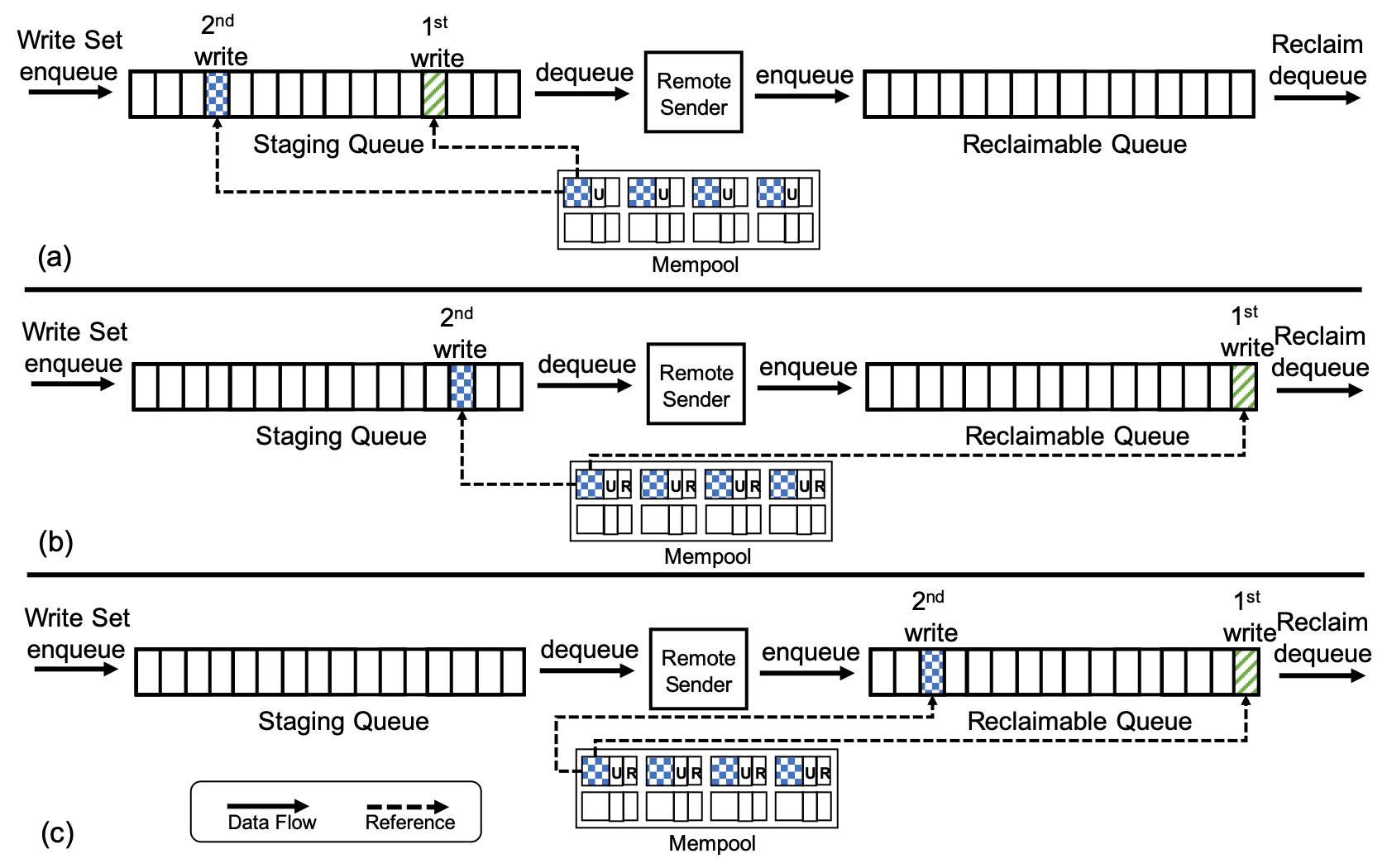}
\caption{Data consistency problem in local mempool and remote replicas due to multiple update requests on the same page. \textmd{ (a) It is solved by a reference counter and an update flag. (b) The case where the distance between two updates are larger than the queue size. (c) The case where the distance is smaller than the queue size.}}
\label{dataconsistency}
\end{figure}

\bigskip
\noindent
\textbf{Problem with multiple updates on the same page.}
There are cases that multiple update write sets are coming on the same page. Then, there are multiple write sets in the Staging queue. The local mempool guarantees the latest data even with the multiple update write sets because the local mempool is always updated immediately and then write sets are enqueued into Staging queue(Figure \ref{dataconsistency} (a)). The problem may occur between remote sending and reclaiming(Figure \ref{dataconsistency} (b)). After 1st write set is sent out and enqueued into Reclaimable queue, 1st write set can be reclaimed before 2nd write is sent out. Then reference pointer for 2nd write set is no longer valid. This is solved by having a simple flag. Each page slot in the work entry has ’Update’ and ‘Reclaimable’ flag. Update flag is set on the pages when the multiple write sets are issued on the same page. When 1st write set is reclaimed, Update flag is examined and skipped. When 2nd is sent out, Update flag is removed from the page slots and page slots will be reclaimed when 2nd write set gets reclaimed. The size of Staging queue and Reclaimable queue is the same. The case that the distance between two write sets is longer than or equal to the queue size can be solved by the Update and Reclaimable flags(Figure \ref{dataconsistency} (b)). Regarding the case that two write sets have shorter distance from each other than the queue size, there is no chance that the 1st write set is reclaimed before 2nd write set is sent out.(Figure \ref{dataconsistency} (c))

\bigskip
\noindent
\textbf{Between replicas and disk.}
Read is always served directly from the local mempool first. Remote node is only accessed when local mempool doesn’t have the page. Likewise, disk is only accessed when remote node fails or pages don't exist in remote nodes including replicas. 
Pages in local mempool can be deleted only when remote sending or disk backup is done and reclaimable flag is set to the pages. Reclaimable pages are tracked by reclaimable queue. If there is an update write set in local mempool and it is not sent out to remote node or disk yet, a reclaimable flag is removed and an update flag is set. The latest page is still served from the mempool until an update write set is sent to remote node or written to the disk.

\subsection{Replication and disk backup}

Valet uses replication as default. Compared to disk writing, replication using RDMA is still faster than writing to disk(Table \ref{latencycomparison}). We use replication for all experiments in evaluation.


\bigskip
\noindent
\textbf{Cost of replication and disk backup.}
With the local mempool, replication and disk backup do not directly add latency to the critical path because replication and disk backup are behind the local mempool and they are out of the critical path. The cost of having replication and/or disk backup approach is memory pressure on MR pool. Slow releasing of unit MR to the MR pool causes shortage of MR in the MR pool and, in turn, it can make getting MR for incoming requests slow too. Another cost of replication is space cost on the remote node. It requires N time larger remote memory space with N replication. 

\begin{figure*}[!htb]
\begin{subfigure}{0.322\textwidth}
\begin{minipage}{0.66\textwidth}
\centering
\includegraphics[width=1\linewidth]{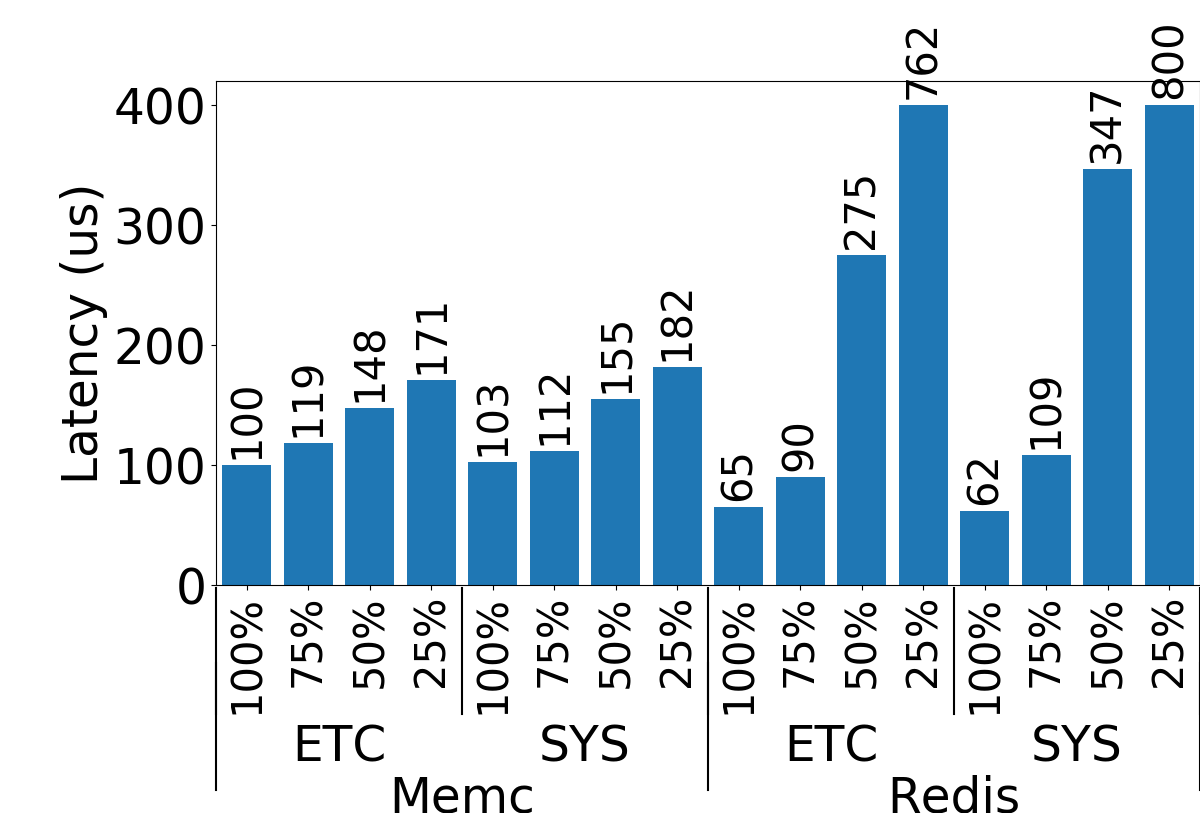}
\end{minipage}
\begin{minipage}{0.3\textwidth}
\centering
\includegraphics[width=1.15\linewidth]{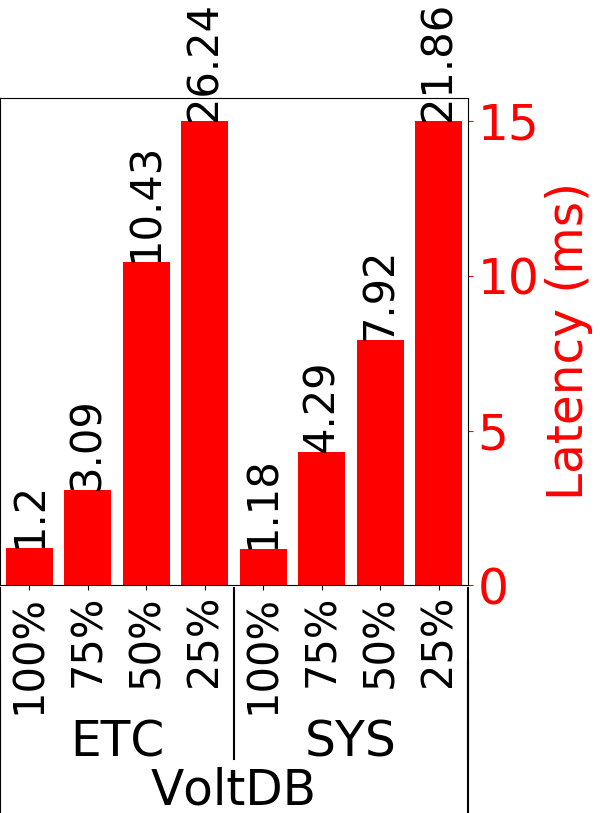}
\end{minipage}\hfill
\caption{nbdX} \label{read_latency:1a}
\end{subfigure}\hspace{0.01\textwidth}
\begin{subfigure}{0.322\textwidth}
\begin{minipage}{0.66\textwidth}
\centering
\includegraphics[width=1\linewidth]{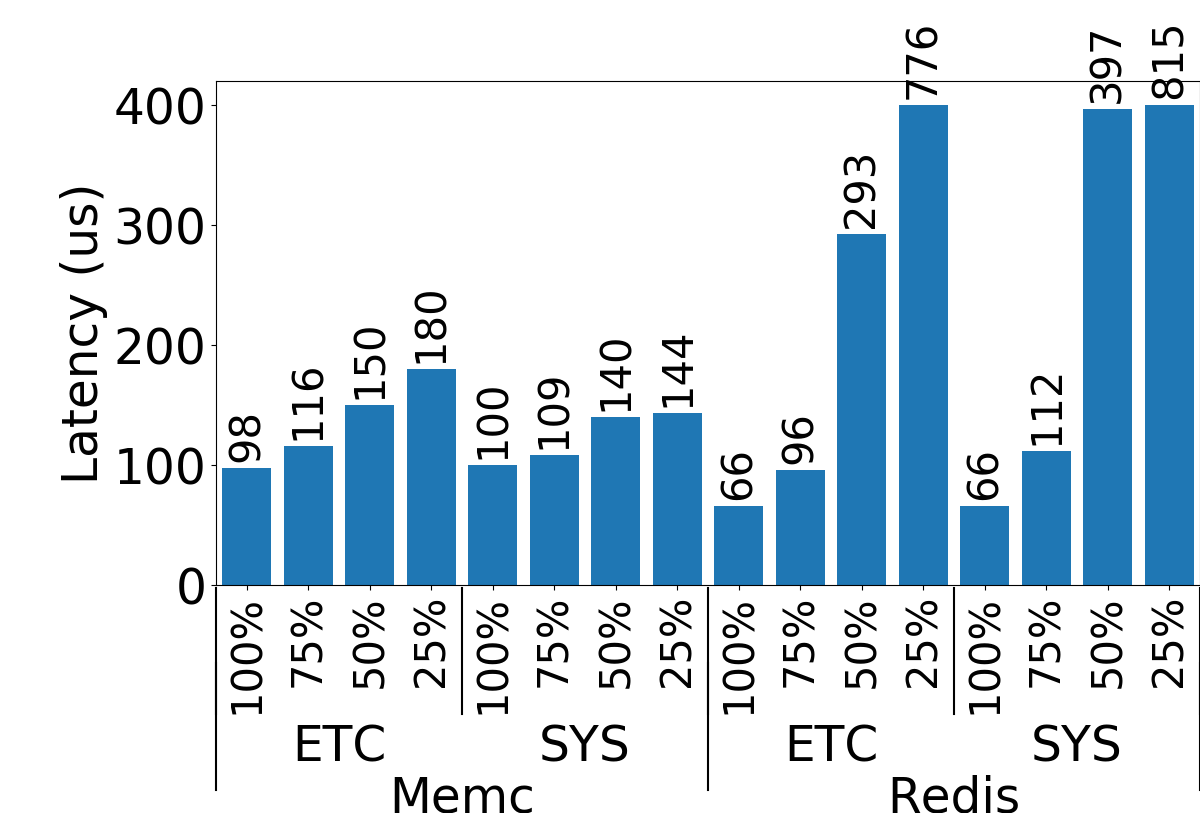}
\end{minipage}
\begin{minipage}{0.3\textwidth} 
\centering
\includegraphics[width=1.15\linewidth]{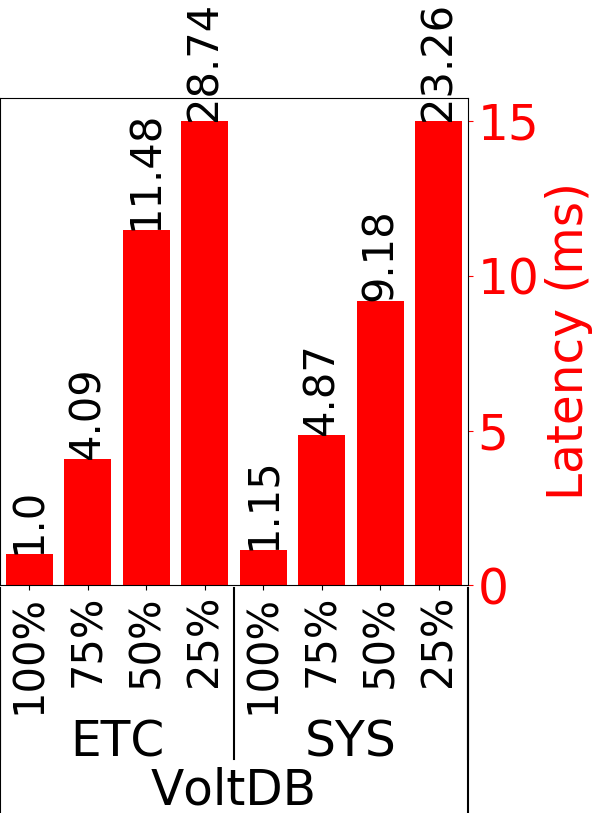}
\end{minipage}\hfill
\caption{Infiniswap} \label{read_latency:1b}
\end{subfigure}\hspace{0.01\textwidth}
\begin{subfigure}{0.322\textwidth}
\begin{minipage}{0.66\textwidth}
\centering
\includegraphics[width=1\linewidth]{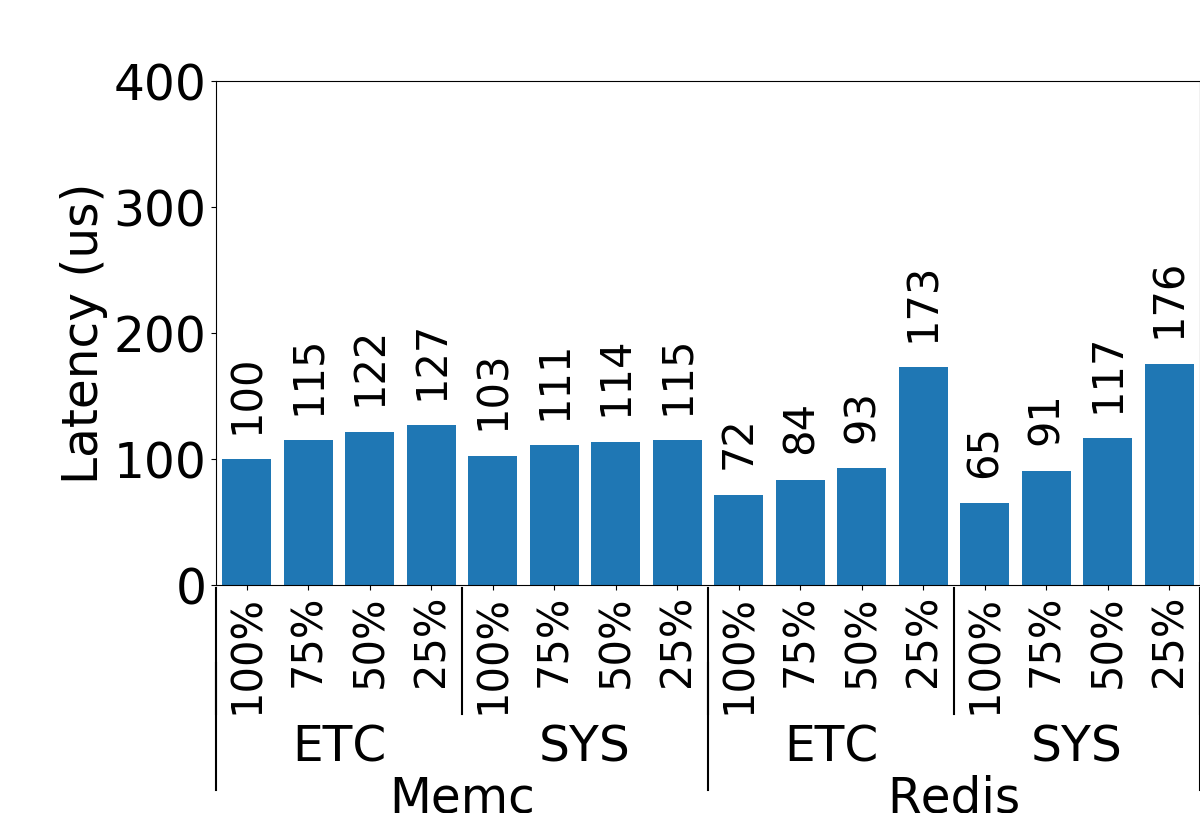}
\end{minipage}
\begin{minipage}{0.3\textwidth}
\centering
\includegraphics[width=1.15\linewidth]{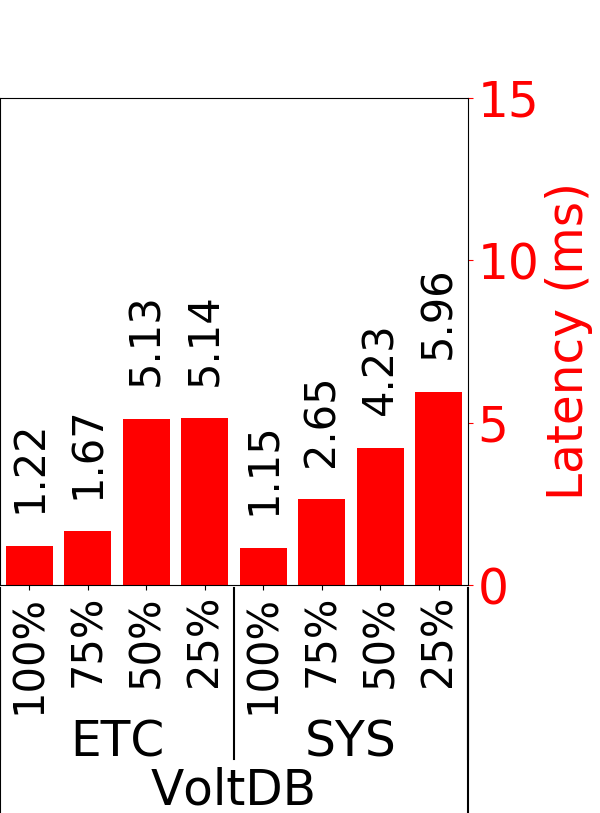}
\end{minipage}\hfill
\caption{Valet} \label{read_latency:1c}
\end{subfigure}
\caption{Big Data Workload Latency Comparison of nbdX, Infiniswap and Valet. (VoltDB scale is in right side)}
\label{read_latency}
\end{figure*}

\begin{figure*}[!htb]
\begin{subfigure}{0.33\textwidth}
\centering
\includegraphics[width=1\linewidth]{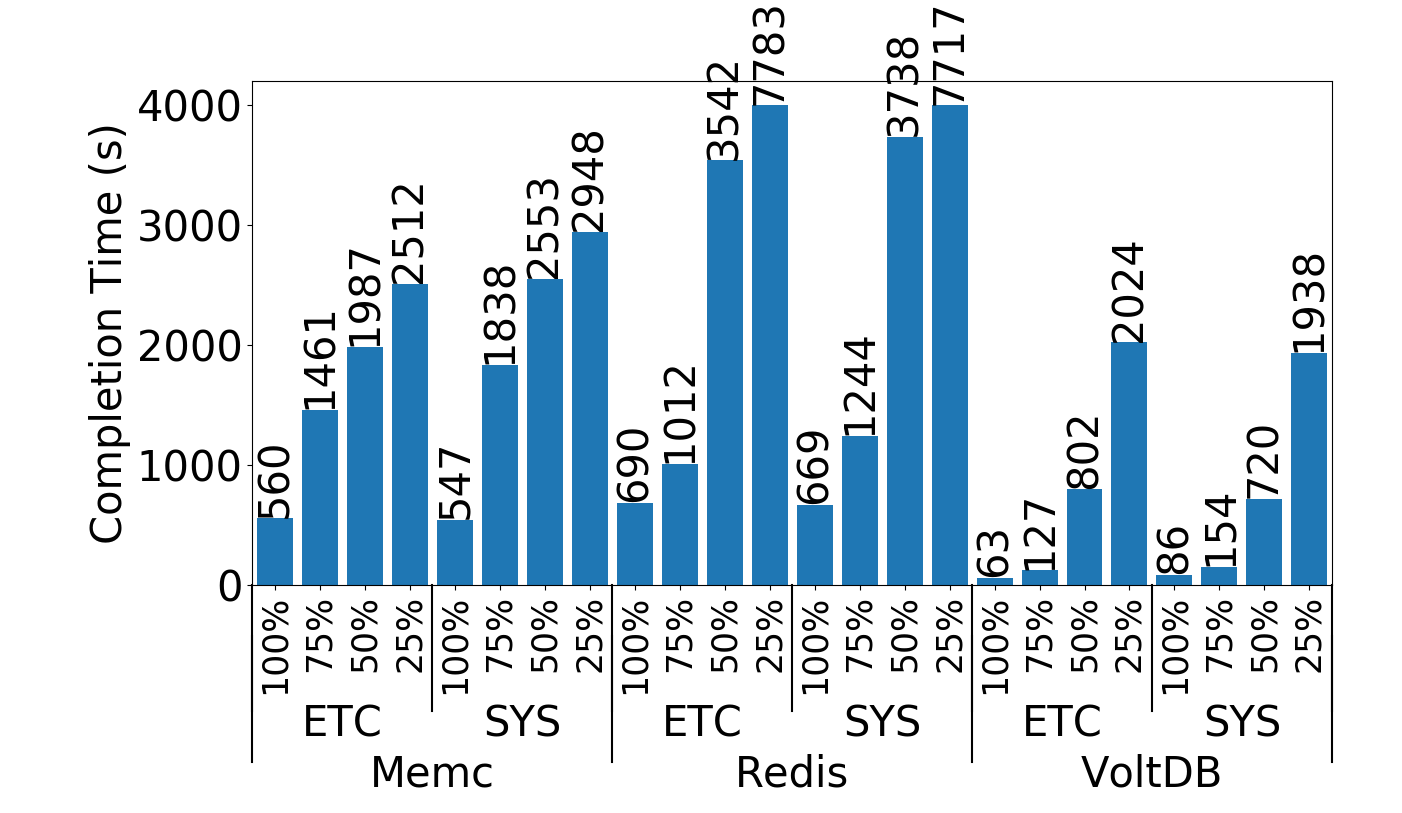}
\caption{nbdX} \label{bd_time:1a}
\end{subfigure}
\begin{subfigure}{0.33\textwidth}
\centering
\includegraphics[width=1\linewidth]{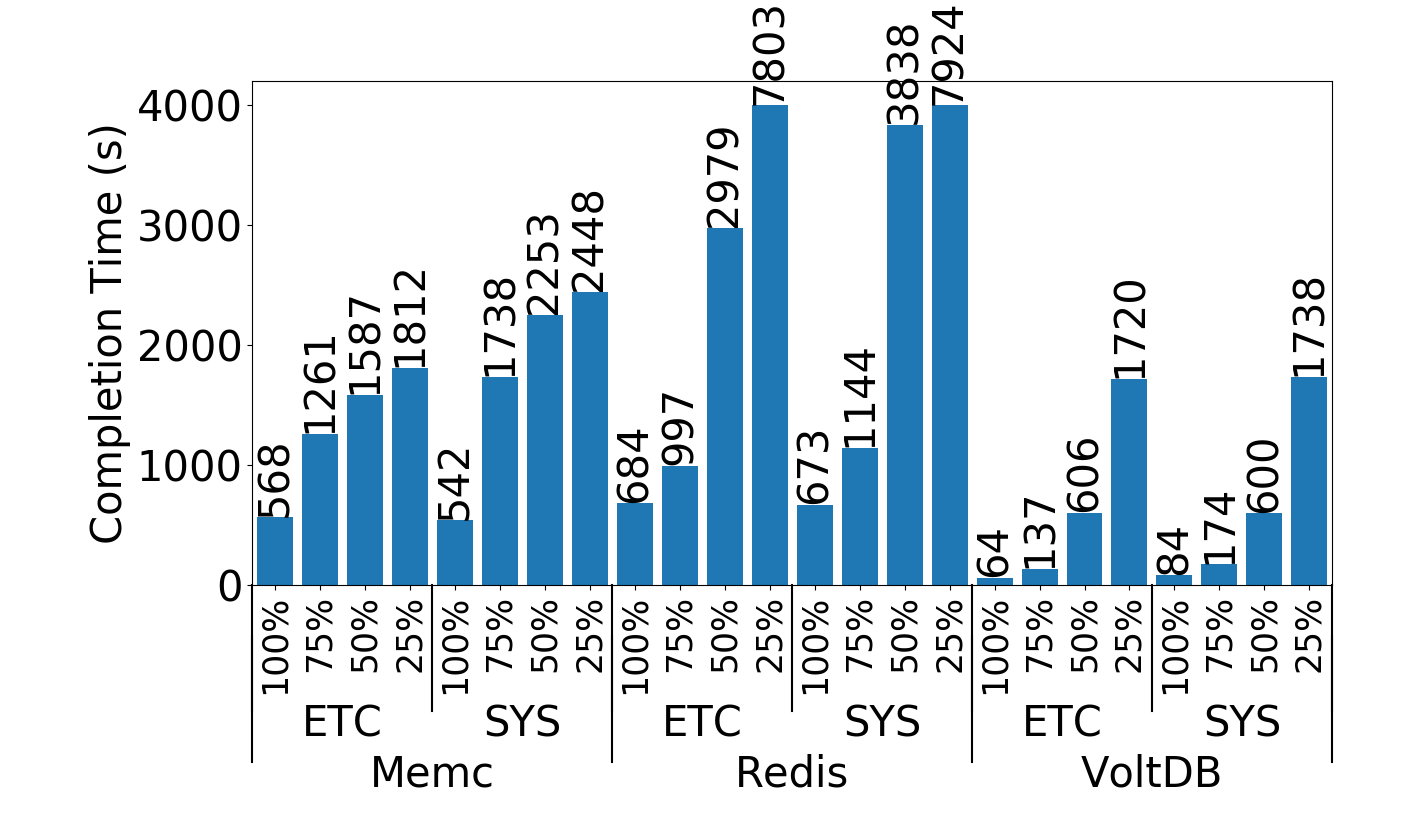}
\caption{Infiniswap} \label{bd_time:1b}
\end{subfigure}
\begin{subfigure}{0.33\textwidth}
\centering
\includegraphics[width=1\linewidth]{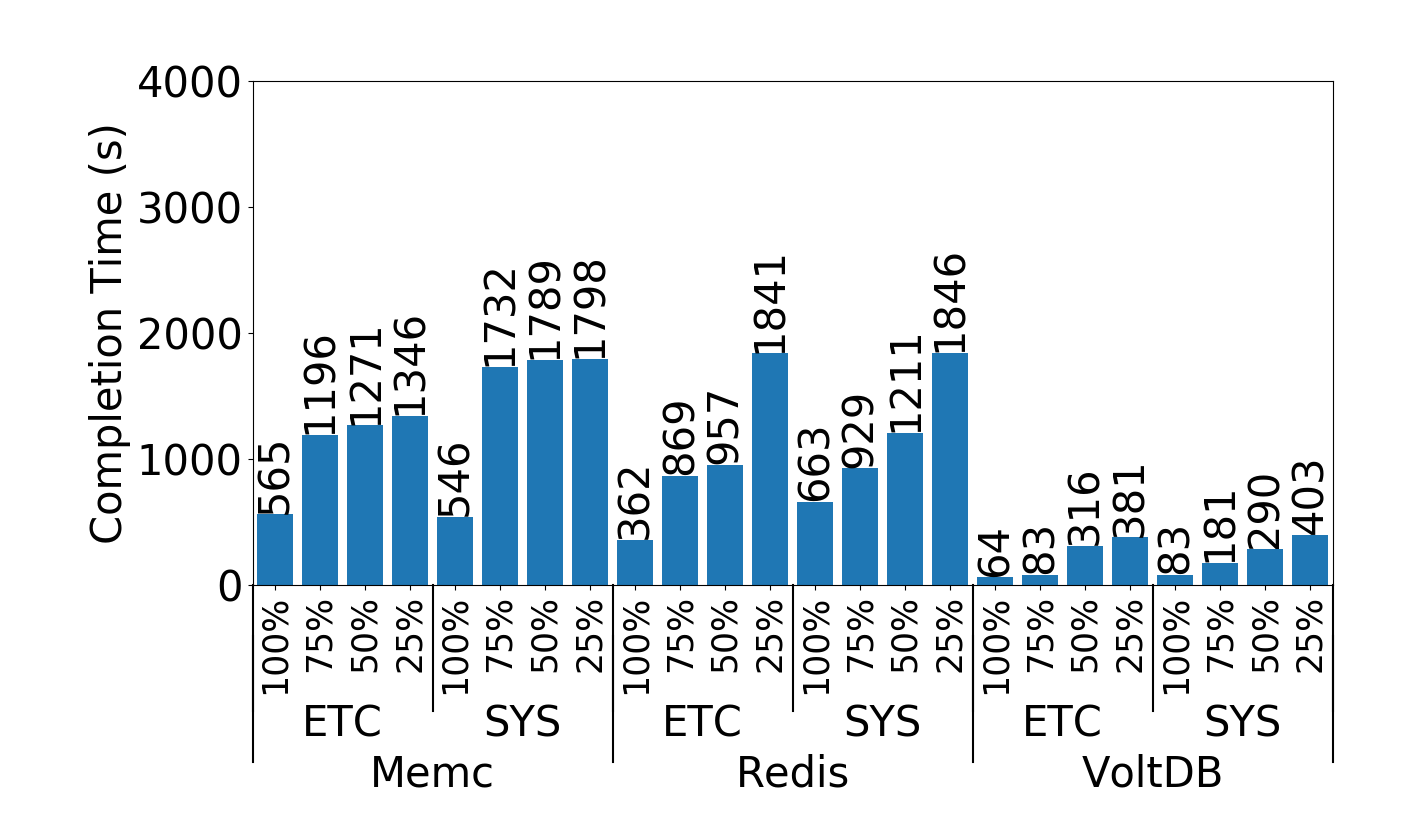}
\caption{Valet} \label{bd_time:1c}
\end{subfigure}
\caption{Big Data workload Completion Time Comparison of nbdX, Infiniswap and Valet}
\label{bd_time}
\end{figure*}

\section{Evaluation}
\label{evaluation}
\textbf{Setup.} We evaluate Valet with eight popular memory intensive applications listed in Table \ref{testdata}. We run five machine learning applications and three big data applications. We run our experiments on \textbf{32 machines with 56Gbps Infiniband cluster} on Cloudlab\cite{Cloudlab}. Each machine has Xeon E5-2650v2 processor(32 virtual cores 2.6Ghz), 64GB memory(DDR-3 1.86Ghz), 1TB SATA 3.5'' rpm hard drives and Mellanox Connect X-3. We run \textbf{90 containers on a 32-machine RDMA cluster} and randomly assign one application on each container. We use 4 different memory limitation on each container. We measure the peak memory usage of each application first. The input dataset sizes are from 10GB to 15GB and these create in-memory working sets from 22GB to 35GB. Each machine has 2 to 3 containers and each container fits workload 100\%, 75\%, 50\% and 25\% in memory respectively. This makes each container to have memory limitation setting from 5GB up to 24GB and paging-out traffic from 5GB to 27GB according to configuration. Unless stated otherwise, we set 64KB block I/O size, 512KB RDMA message size and 1GB as an unit size of remote MR block. The size of the local mempool will be specified in each experiment. For stable measurement, average is taken from 5 times run for each case. We compare our system with Infiniswap\cite{Juncheng} and nbdX\cite{nbdX}. We set Infiniswap as default as their paper mentions and nbdX uses remote ramdisk for storing data. 

\begin{table}[ht!]
\centering
\includegraphics[width=80mm]{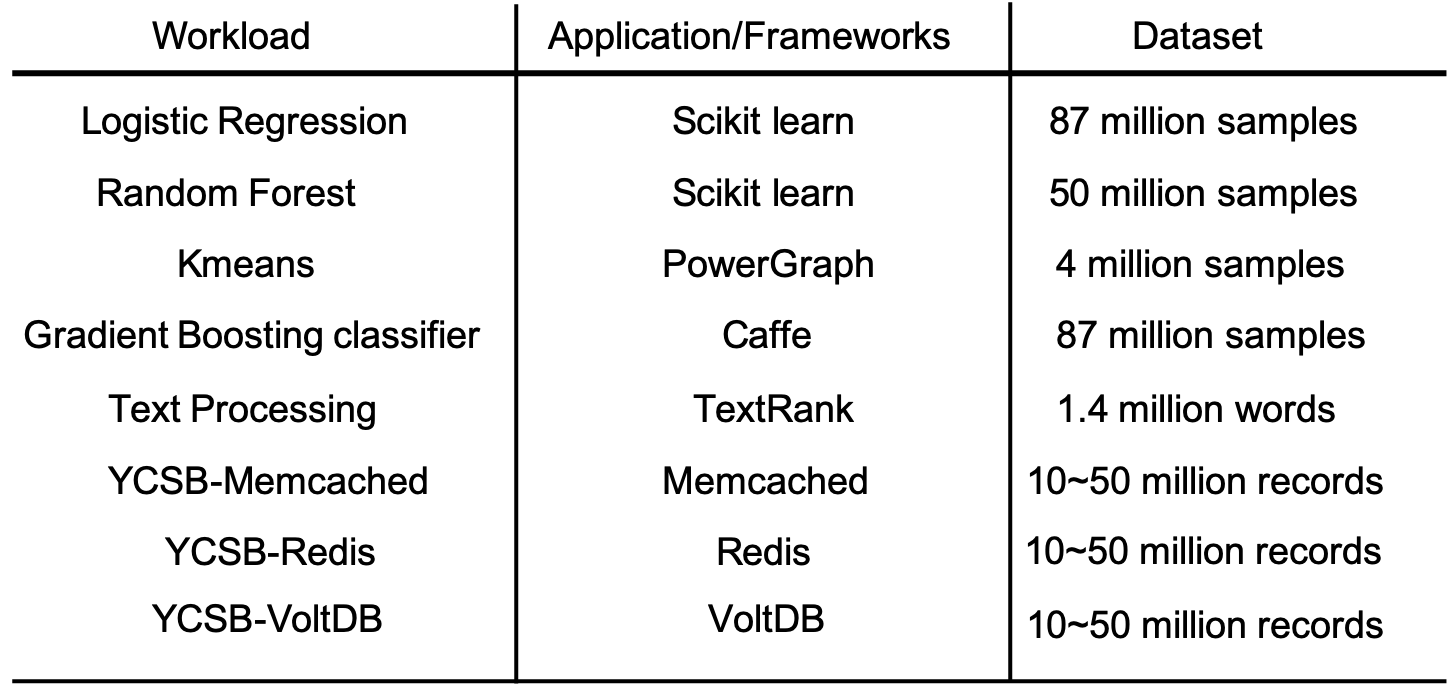}
\caption{ Applications and workload used in evaluation.}
\label{testdata}
\end{table}








\subsection{BigData Workload Performance}

In this experiment, we measure Valet performance on Memcached\cite{Memcached}, Redis\cite{Redis} and VoltDB\cite{VoltDB}. Memcached and Redis is in-memory distributed caching system through simple key-value interface. VoltDB is ACID-compliant in-memory transactional database. We compare Valet(Figure \ref{bd_time:1c}) to Infiniswap(Figure \ref{bd_time:1b}) and nbdX(Figure \ref{bd_time:1a}). For workload, we use Facebook simulated workload\cite{Facebookworkload} ETC and SYS by using YCSB\cite{YCSB}. We use zipfian distribution for both workload. We first populate the applications with 10 million records in advance and run 10 million queries with ETC and SYS workload. Dataset size is 10GB and working set memory with this dataset ranges from 15GB to 22GB. Each application takes different amount of working set memory after we populate and run the same 10GB workload. Peak memory for Memcached is 15GB and 22GB for both Redis and VoltDB. Compared to simple key-value structure such as Memcached, its complicated data structure in VoltDB requires more memory. For local mempool setting, we set local mempool dynamically expands and shrinks based on free memory on the host node.

First, Valet shows more stable performance than Infiniswap and nbdX. See Figure \ref{bd_time}. nbdX and Infiniswap's completion time increases superlinearly as more pages are sent to remote nodes whereas Valet shows steady performance. Table \ref{perfgapbd} shows summary of performance improvement comparison in Figure \ref{bd_time}. Valet outperforms up to 5.3$\times$\ over nbdX and up to 4.5$\times$\ over Infiniswap. Valet also outperformas Linux conventional OS swap by up to 128$\times$. Note that we didn't put the figure of completion time of Linux conventional OS swap.

\begin{table}[t]
\centering
\begin{tabular}{ |p{2cm}||p{1.6cm}|p{1.6cm}|p{1.6cm}| }
\hline
\multicolumn{4}{|c|}{Valet's improvement over other systems (BigData)} \\
\hline
\centering
WorkingSet Fit&Linux &nbdX&Infiniswap\\
\hline
\centering
75\% &41x(86x)&1.19x(1.5x)&1.17x(1.7x)\\
\centering
50\% &59x(101x)&2.5x(3.7x)&2.1x(3.2x)\\
\centering
25\% &90x(138x)&3.7x(5.3x)&3.3x(4.5x)\\
\hline
\end{tabular}
\caption{Summary of performance improvement comparison of Valet with other systems in Figure \ref{bd_time} and Linux. \textmd{It show improvement on average and on best case in brackets.}}
\label{perfgapbd}
\end{table}

Second, the performance gap between Valet and other systems increases as more pages are sent to remote nodes. See Table \ref{perfgapbd}. nbdX and Infiniswap's perofrmance is not scalable well compared to Valet as percentage of working set fit decreases.

Third, we also measure average latency of each application on three systems(Figure \ref{read_latency}). Compared to 100\% working set in-memory fit case, Valet latency increases 1.5$\times$, 2.4$\times$ and 3$\times$ in 75\%, 50\% and 25\% fit case respectively. nbdX latency increases 2$\times$, 4.9$\times$ and 12$\times$ in 75\%, 50\% and 25\% fit case respectively. Infiniswap latency increases 2.3$\times$, 5.5$\times$ and 12.7$\times$ in 75\%, 50\% and 25\% fit case respectively. Conventional OS swap facility latency increases 39.8$\times$, 58.8$\times$ and 93.4$\times$ in 75\%, 50\% and 25\% fit case respectively.

\begin{figure*}[!htb]
\begin{subfigure}{0.33\textwidth}
\centering
\includegraphics[width=1\linewidth]{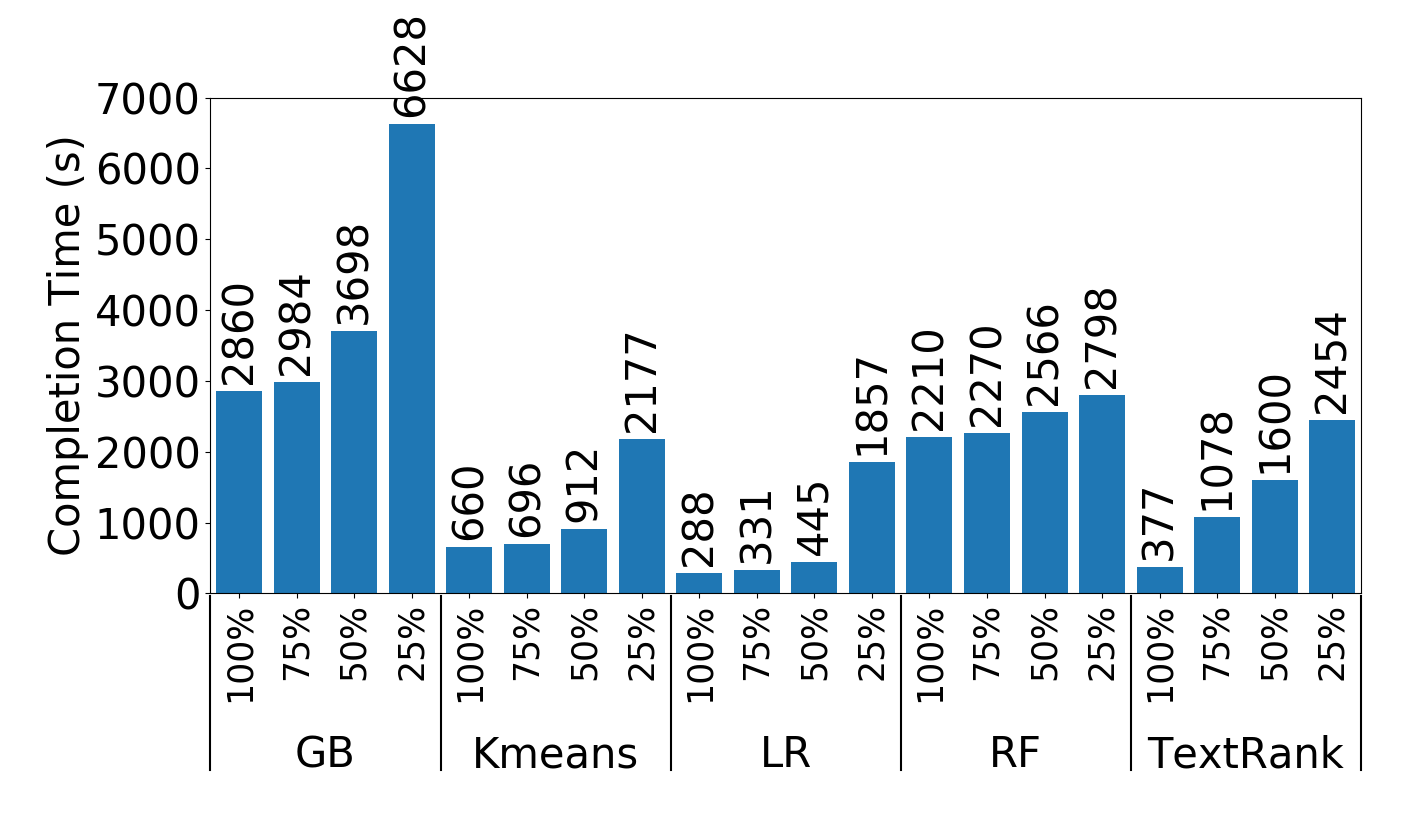}
\caption{nbdX} \label{ml_time:1a}
\end{subfigure}
\begin{subfigure}{0.33\textwidth}
\centering
\includegraphics[width=1\linewidth]{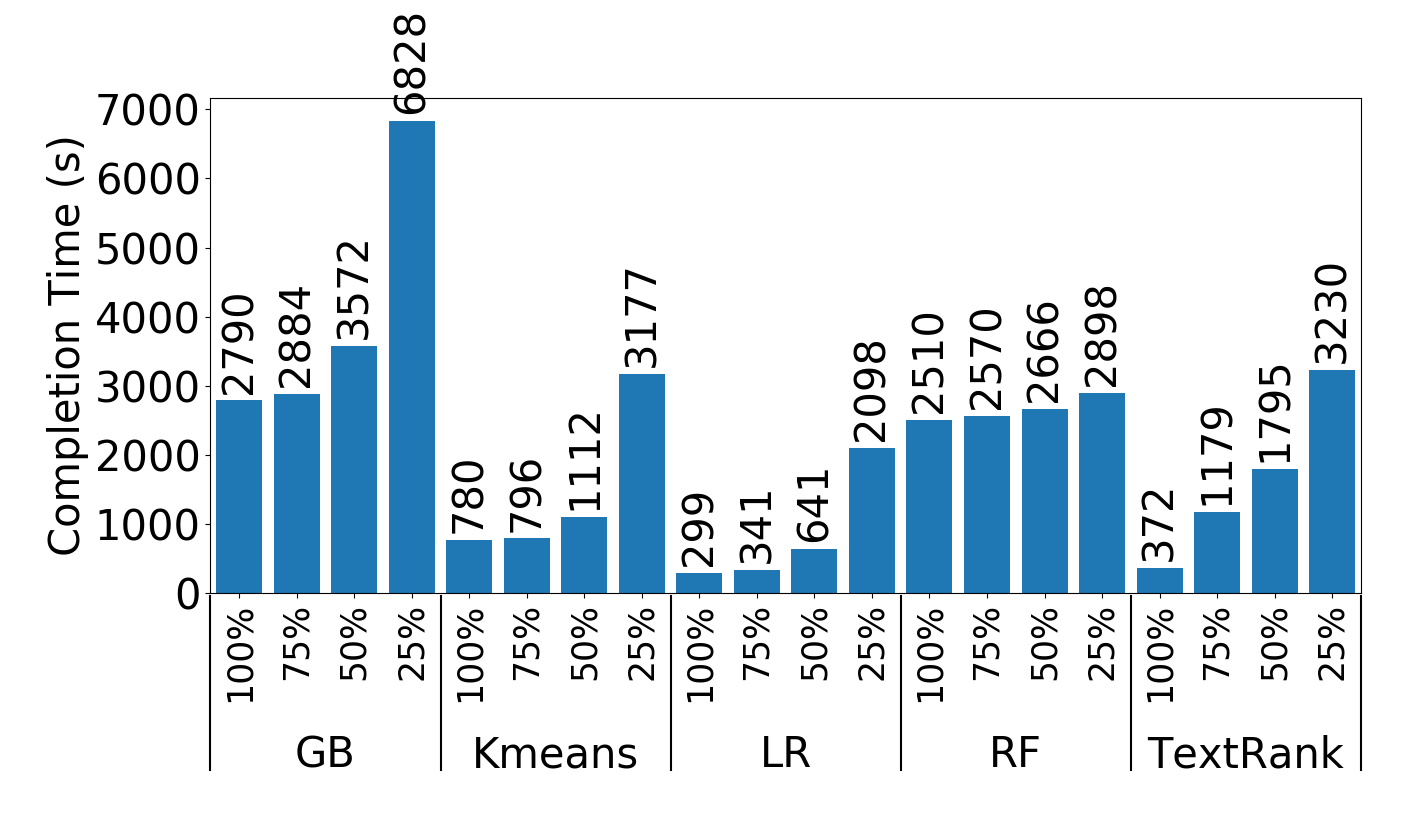}
\caption{Infiniswap} \label{ml_time:1b}
\end{subfigure}
\begin{subfigure}{0.33\textwidth}
\centering
\includegraphics[width=1\linewidth]{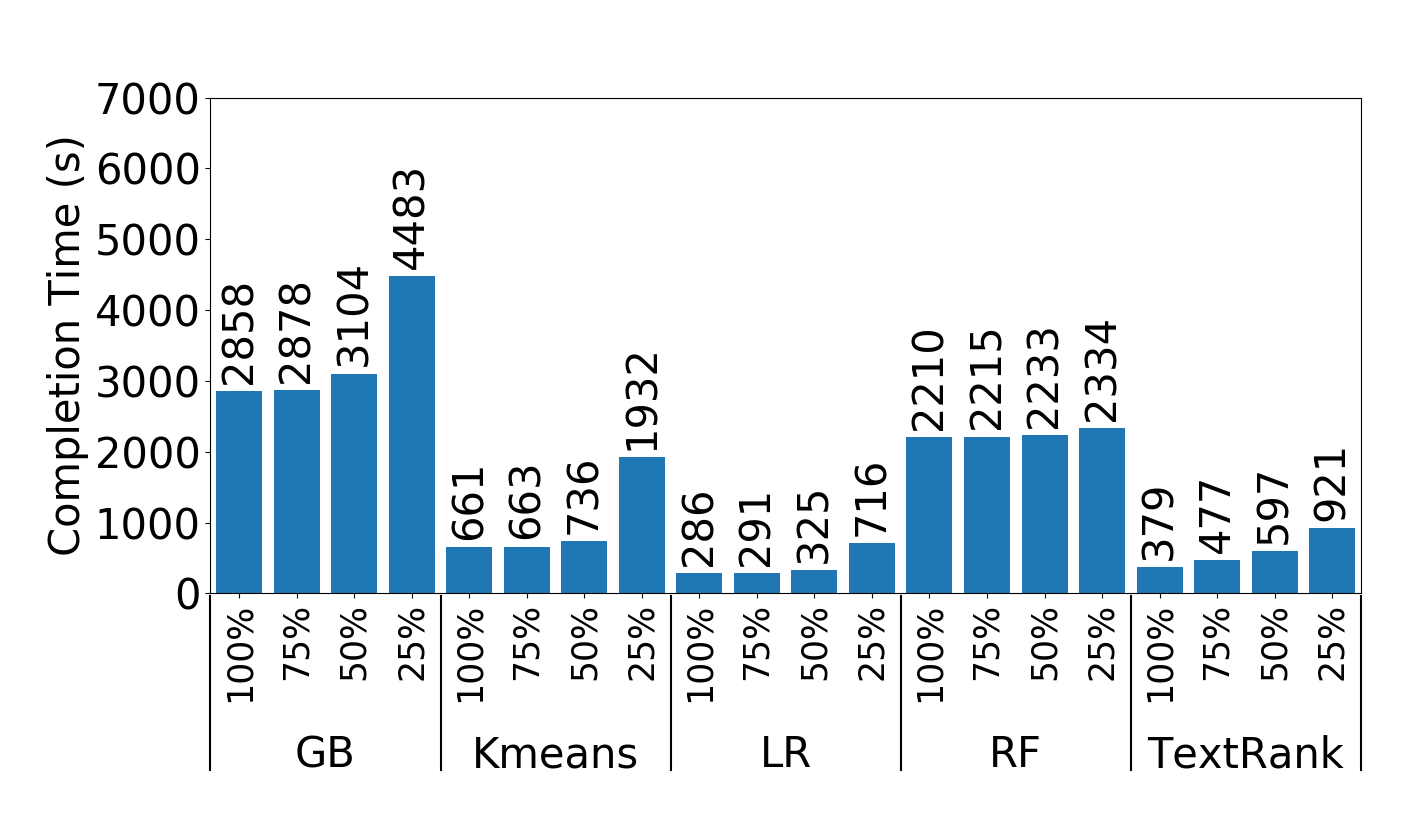}
\caption{Valet} \label{ml_time:1c}
\end{subfigure}
\caption{Machine Learning workload Completion Time Comparison of nbdX, Infiniswap and Valet}
\label{ml_time}
\end{figure*}


\begin{table}[t]
\centering
\begin{tabular}{ |p{2cm}||p{1.6cm}|p{1.6cm}|p{1.6cm}| }
\hline
\multicolumn{4}{|c|}{Valet's improvement over other systems (ML)} \\
\hline
\centering
WorkingSet Fit& Linux &nbdX&Infiniswap\\
\hline
\centering
75\% &47x(99x)&1.3x(2.3x)&1.4x(2.5x)\\
\centering
50\% &69x(187x)&1.5x(2.7x)&1.8x(3x)\\
\centering
25\% &109x(282x)&1.8x(2.7x)&2.2x(3.5x)\\
\hline
\end{tabular}
\caption{Summary of performance improvement comparison of Valet with other systems in Figure \ref{ml_time} and Linux. \textmd{It show improvement on average and on best case in brackets.}}
\label{perfgapml}
\end{table}

\bigskip
\noindent
\subsection{ML Workload Performance}

We use various popular Machine Learning workload(Gradient Boosting classifier, Kmeans, Logistic Regression, Random Forest and TextRank) to measure performance of Valet and other systems in Figure \ref{ml_time}. Datasets we use are from 4 million to 87 million samples and they create from 9GB to 34GB workload. For ML, we use click prediction data from Kaggle\cite{mldata1} and NOAA weather dataset\cite{mldata2}. For TextRank, we use wiki dataset \cite{textrankdata}, which includes 1.4 million words. We also apply 75\%, 50\% and 25\% working set fit. For local mempool setting, we also set local mempool dynamically expands and shrinks based on free memory on the host node.

Table \ref{perfgapml} shows summary of performance improvement comparison in Figure \ref{ml_time}. Valet outperforms by up to 282$\times$ over Linux, by up to 2.7$\times$\ over nbdX and by up to 3.5$\times$\ over Infiniswap. Valet generally shows more stable performance than Infiniswap and nbdX like BigData workload. An interesting observation is that nbdX's and Infiniswap's completion time increases superlinearly as workload increases except Kmean. We observed that Kmean's access pattern is different from others. It intensively accesses certain MR blocks that are mapped in early stage of running rather than access various MR blocks. Since those intensive accessing memory blocks are assigned in early stage of running, it is highly likely in-memory in the local host. This repetitive access pattern also might increase page cache hit in OS. For now, Valet uses LRU on local mempool. Cache replacement policy like MRU that works well on repetitive access pattern might be useful for local mempool replacement policy for this type of workload. We leave this exploration as a future work. 

\subsection{Effectiveness of optimization}
\begin{figure}[H]
\begin{subfigure}{0.49\textwidth}
\centering
\includegraphics[width=80mm,height=20mm]{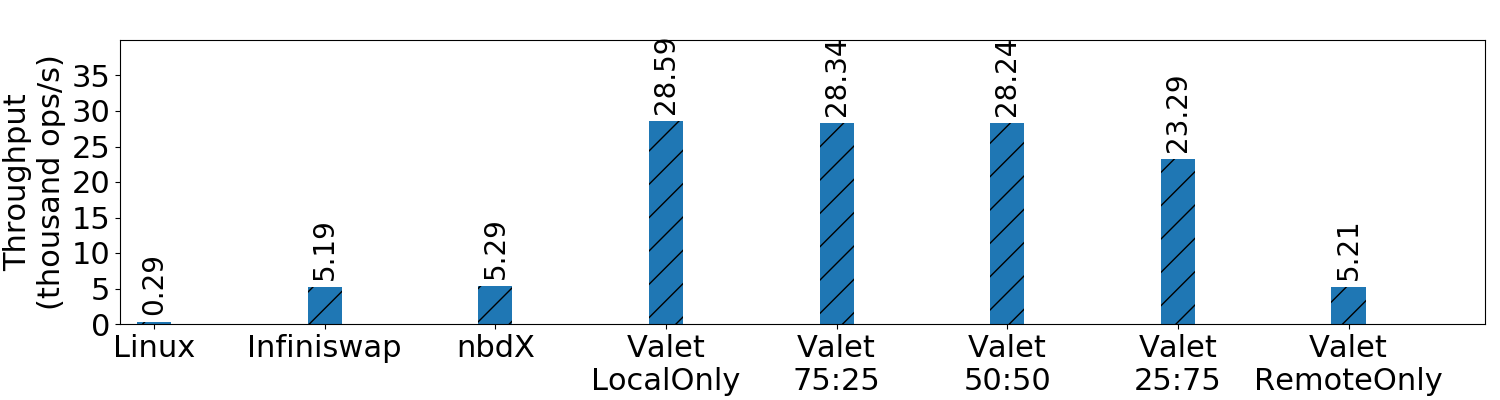}
\caption{VoltDB} \label{mempool_eval:1a}
\end{subfigure}\vfill
\begin{subfigure}{0.49\textwidth}
\centering
\includegraphics[width=80mm,height=20mm]{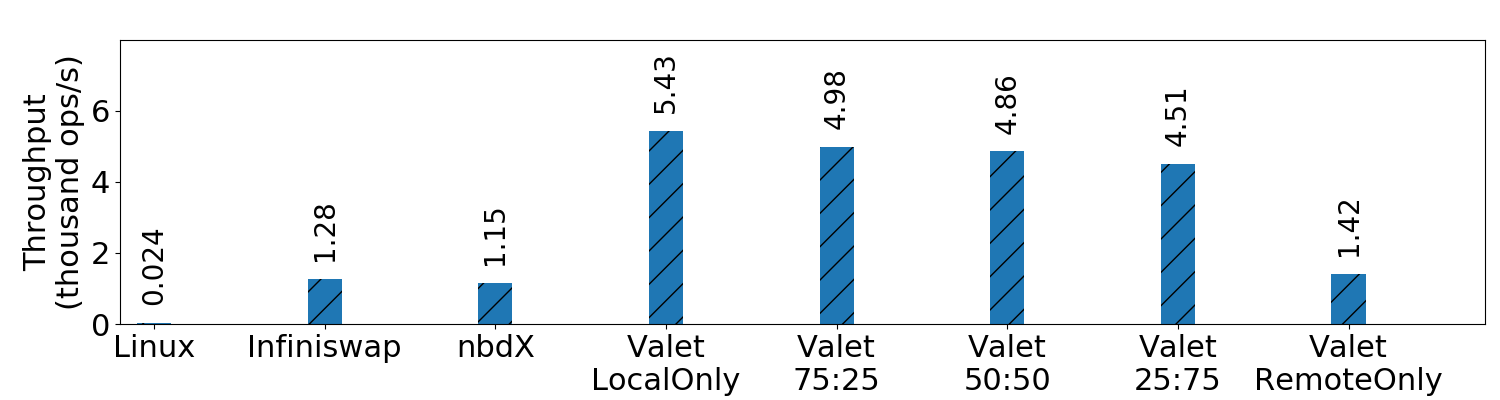}
\caption{Redis} \label{mempool_eval:1b}
\end{subfigure}\vfill
\begin{subfigure}{0.49\textwidth}
\centering
\includegraphics[width=80mm,height=20mm]{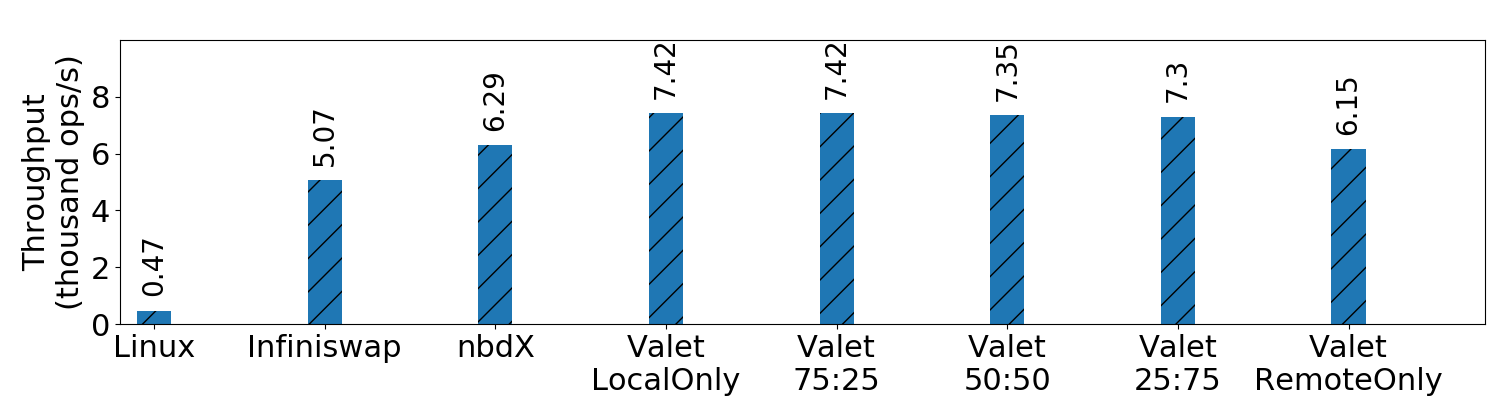}
\caption{Memcached} \label{mempool_eval:1c}
\end{subfigure}\vfill
\caption{Impact of Host/Remote memory distribution}
\label{mempool_eval}
\end{figure}

\textbf{Host/Remote memory distribution}
This section compares performance impact of various host/remote memory ratio on application for conventional OS swap(Linux), nbdX, Infiniswap and Valet. We use 25\% working set fit configuration for all four systems(Linux, nbdX,Infiniswap and Valet). 75\% of working set workload is distributed across remote nodes via paging. For Valet, Valet-75:25, Valet-50:50 and Valet-25:75 denote ratio of local memory to remote memory working set. Valet-LocalOnly and Valet-RemoteOnly denote all working set resides in local node and remote node respectively. Figure \ref{mempool_eval} shows the comparison. 

We highlight several observations below. First, using Valet-LocalOnly, throughput of VoltDB, Redis and Memcached increase by up to 98.5$\times$\, 226.26$\times$\ and 15.7$\times$\ compared to Linux, increase by up to 5.5$\times$\, 3$\times$\ and 1.46$\times$\ compared to Infiniswap, and increase by up to 5.4$\times$\, 4.7$\times$\ and 1.17$\times$\ compared to nbdX. 

Second, throughput increases as the size of local mempool increases from Valet-RemoteOnly to Valet-LocalOnly. However, the performance gap between Valet-RemoteOnly and Valet-25:75 is the largest when increasing the size of mempool. Note that Valet-RemoteOnly does not have local mempool component. It shows that critical path optimization with local mempool is the most effective improvement in this experiment. 

Third, even with Valet-25:75 that fits only 25\% of workload in memory, its performance is comparable to larger percentage cases. By pipelining local mempool in the critical path, it effectively reduces latency(\cref{latencyoverheadsolution}). Pages in the mempool are sent to remote and replaced by newly incoming pages. Bigger sized mempool gets more pages in the mempool and it can provide higher local cache hit and, in turn, provide more performance.

\noindent
\textbf{Critical path optimization impact on latency}
In Table \ref{evictionimpact}, we measure latencies of every events in the critical path with Valet-25:75 setting in Valet and Infiniswap. For workload, we use VoltDB with YCSB SYS workload. ETC and SYS are Facebook simulated workload\cite{Facebookworkload}. ETC is read heavy workload that contains 95\% of GET and 5\% of SET. SYS is write heavy workload that contains 75\% of GET and 25\% of SET. In this measurement Valet enables Disk Backup for fair comparison with Infiniswap. Disk access happens when data is not found on remote node(e.g. remote eviction) or there is no connection to node or mapping to MR block. As we expected, Infiniswap's latency is severely affected by disk access(\ref{evictionimpact:1b}). Infiniswap redirects request traffic to disk while connection and mapping is setup. Valet, on the other hand, avoids disk access due to connection or mapping by having local mempool in the critical path(\ref{evictionimpact:1a}). Request traffic goes to local mempool first and is sent to remote node later. 25\% local hit helps to lower the latency further in read request. Write request only spends latency regarding local storing, which is radix tree insertion to track the pages in the local node, data copy from BlO structure to local mempool and enqueueing request to staging queue to track remote sending. Write request doesn't wait RDMA sending part unlike Infiniswap. Latencies for connection, mapping and disk access are also hidden from critical path. 
Although connection and mapping are also hidden from write critical path in Infiniswap, the delay causes disk access and, in turn, disk access is not hidden from critical path.

\begin{table}[H]
\begin{subfigure}{0.49\textwidth}
\centering
\includegraphics[width=1\linewidth]{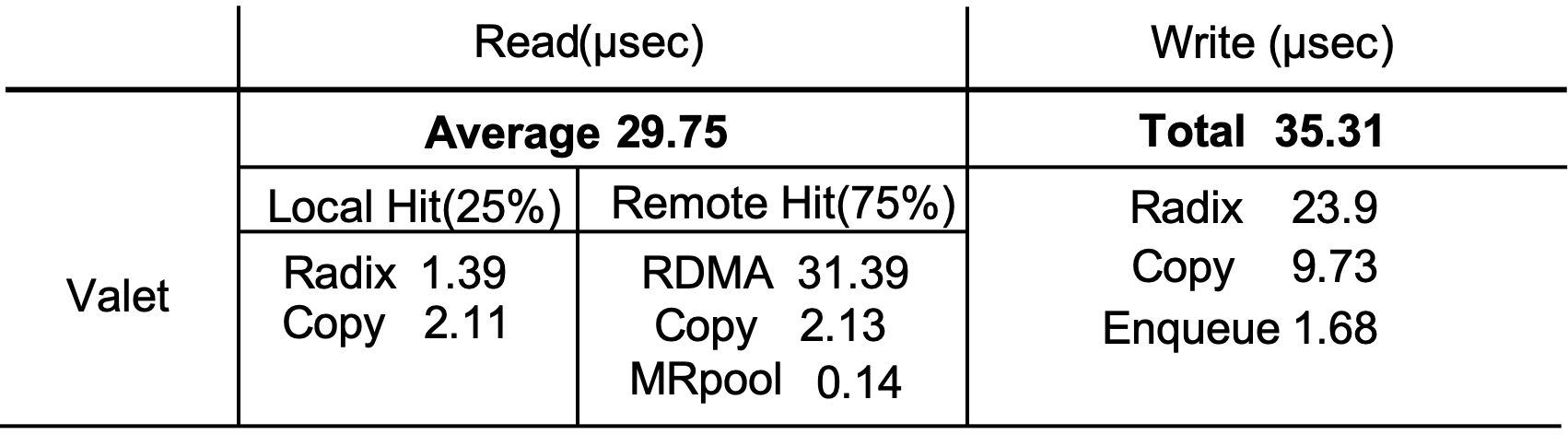}
\caption{Valet} \label{evictionimpact:1a}
\end{subfigure}\vfill
\begin{subfigure}{0.49\textwidth}
\centering
\includegraphics[width=1\linewidth]{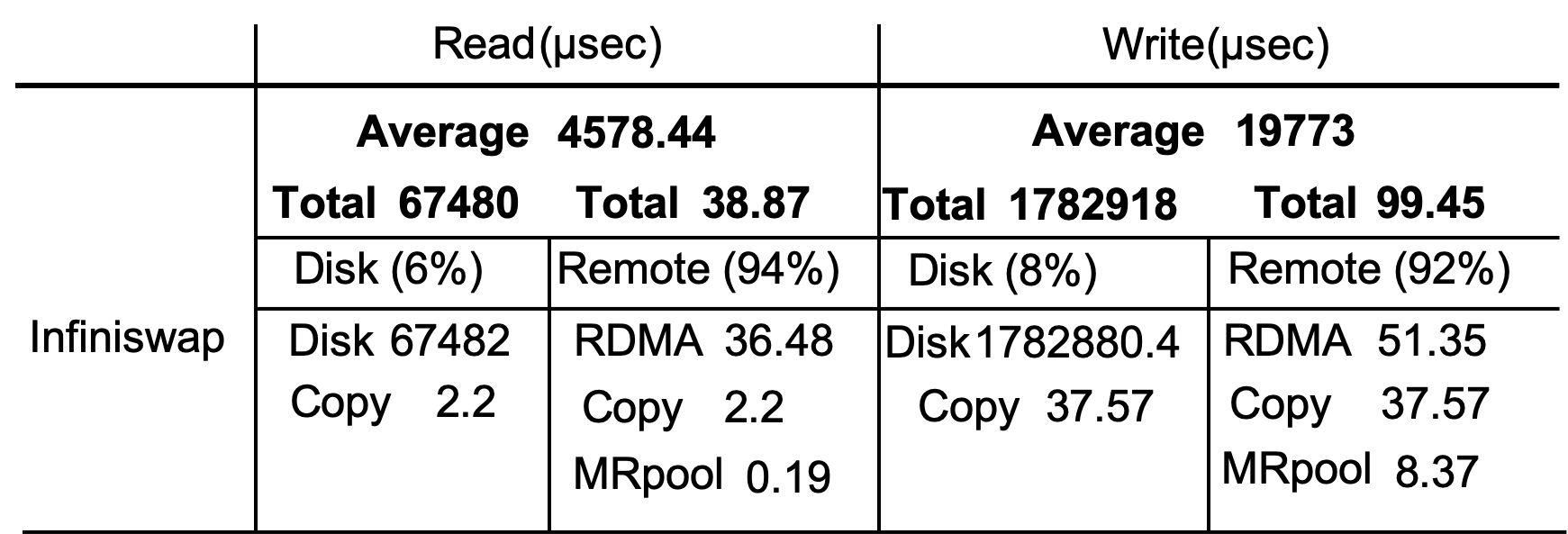}
\caption{Infiniswap} \label{evictionimpact:1b}
\end{subfigure}\vfill
\caption{latency breakdown comparison between Valet and Infiniswap. }
\label{evictionimpact}
\end{table}



\subsection{Scalability}

\begin{figure}[!htb]
\begin{minipage}{0.25\textwidth}
\centering
\includegraphics[width=1\linewidth]{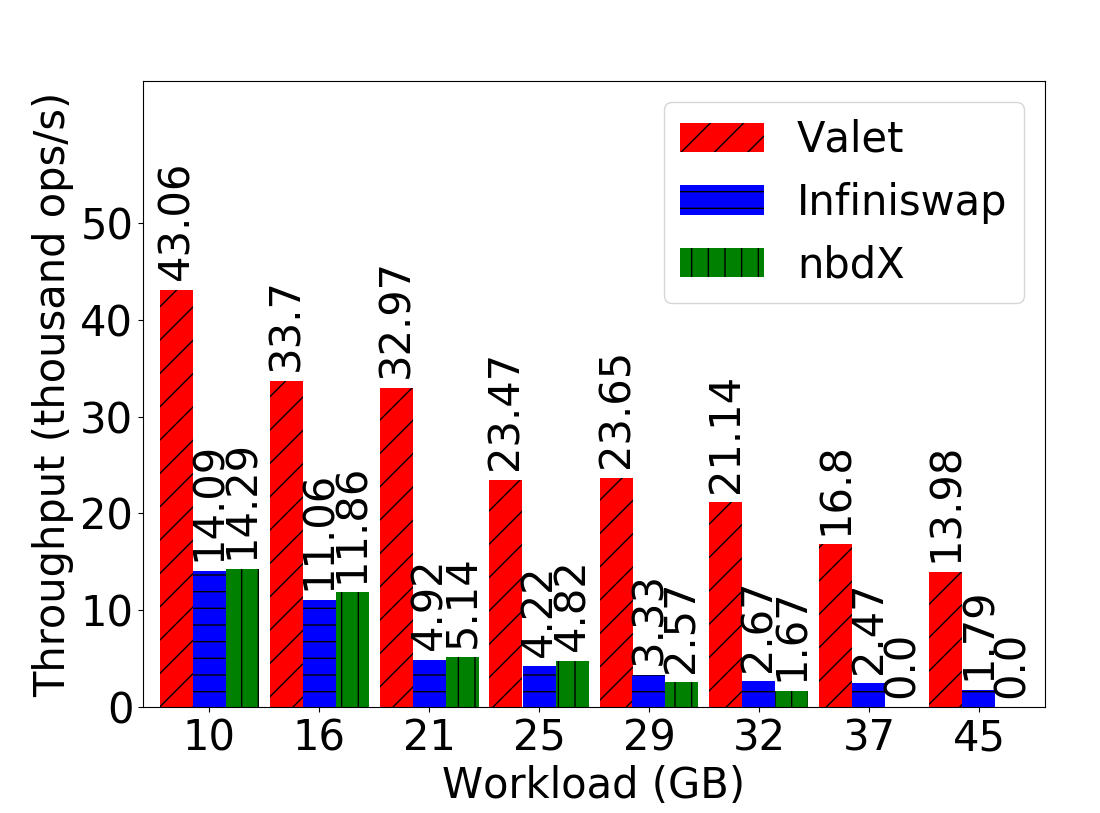}
\end{minipage}\hfill
\begin {minipage}{0.22\textwidth}
\centering
\includegraphics[width=1.1\linewidth,height=34mm]{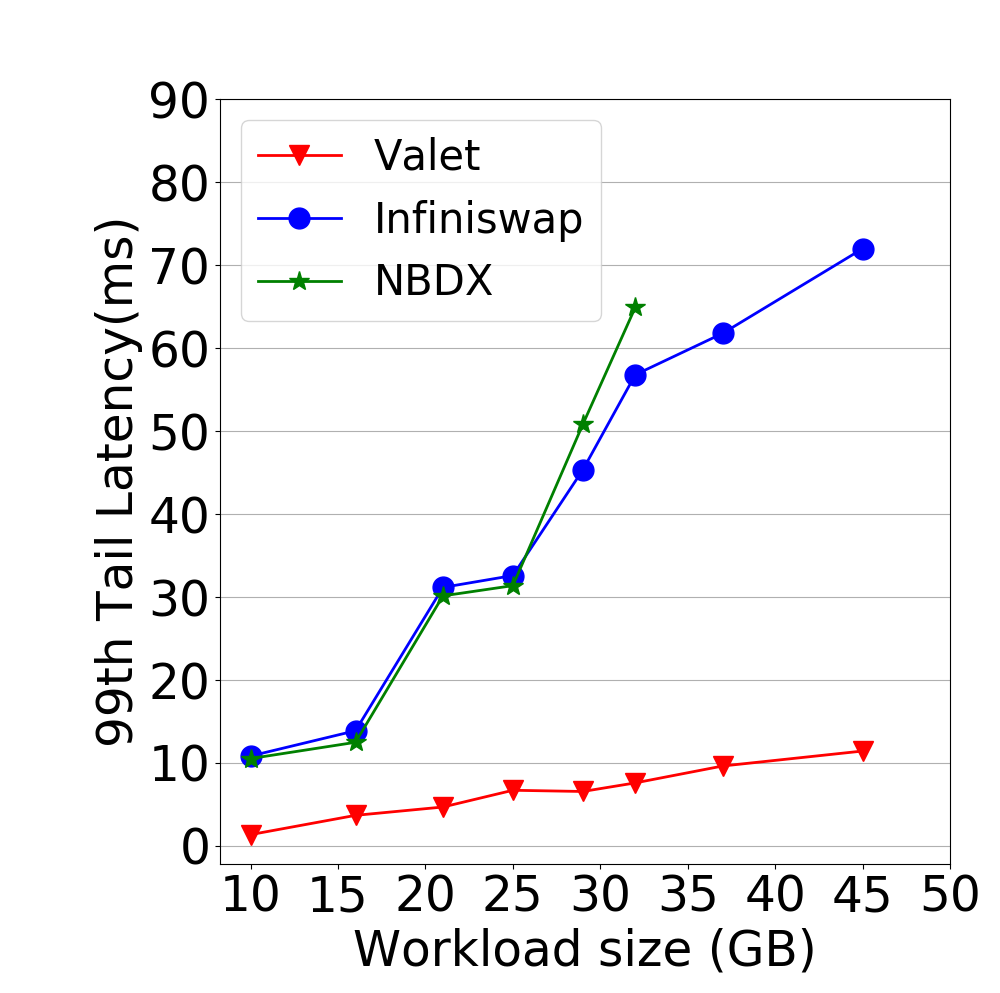}
\end{minipage}
\caption{Scalability comparison between Valet, nbdX and Infiniswap with increasing workload. }
\label{scalability_test}
\end{figure}

In paging system, it is important that sender node handles increasing workload well. In this experiment, we try to figure out Valet's effectiveness with large workload and scalability(Figure \ref{scalability_test}). We choose VoltDB because it has the poorest latency among other applications. we measure throughput and 99th percentile tail latency. For Valet, we use 500MB fixed size local mempool to avoid the benefit of the local memory but to include the benefit of critical path optimization. Throughput decreases as workload increases but Valet still outperforms by up to 7.8$\times$\ over Infiniswap and by up to 12.65$\times$\ over nbdX in throughput. 99th percentile tail latency increases by up to 6.45$\times$\ in Infiniswap and by up to 7.2$\times$\ in nbdX over Valet. Note that we were not able to measure nbdX with larger workload than 32GB due to unstable running. nbdX uses two sided verb with message pool on both sender and receiver node. We observe sender and receiver side message pool becomes the bottleneck and it severely drops the performance during this experiment.

\subsection{Eviction Cost}

\begin{figure}[ht!]
\centering
\includegraphics[width=80mm]{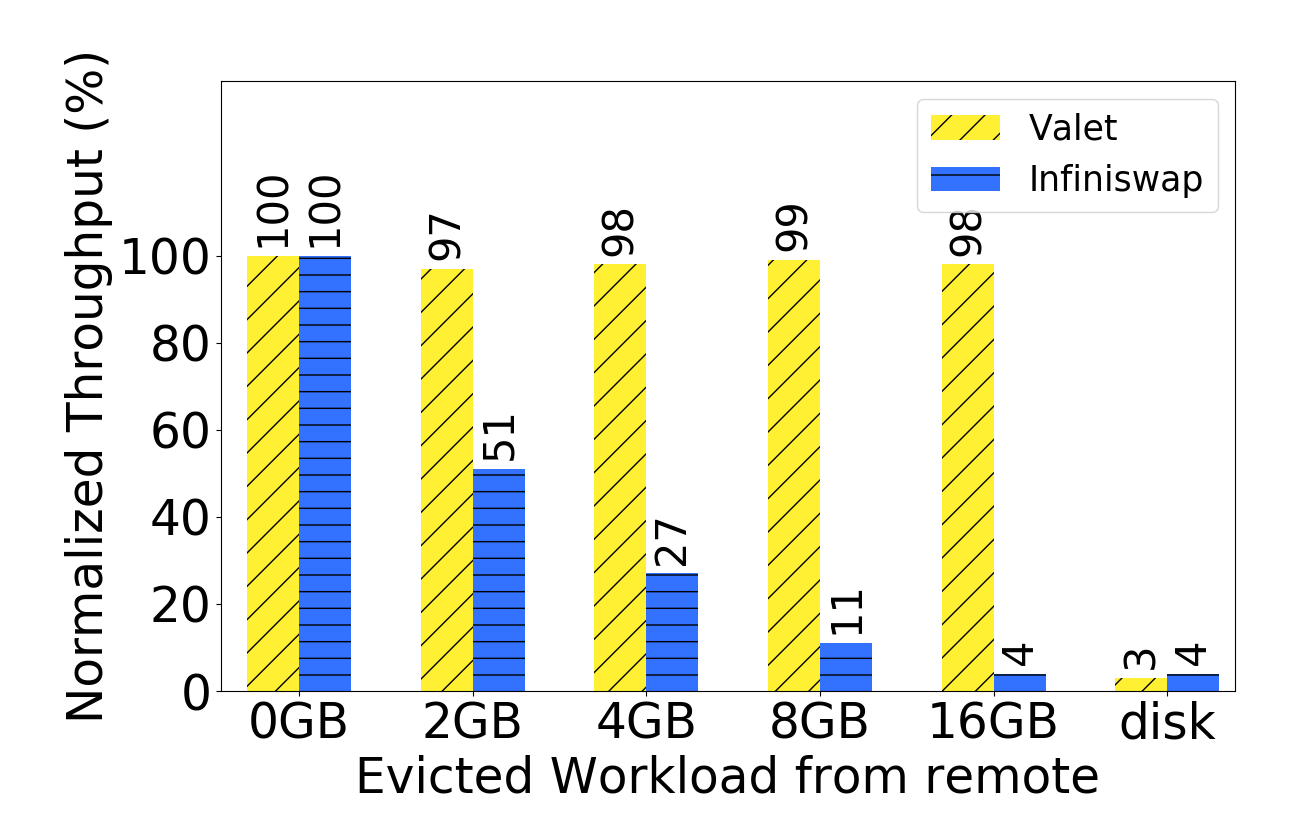}
\caption{No remote eviction impact in Valet by migration instead of eviction. \textmd{We run Redis with 20GB workload. About 16GB is distributed into remote nodes.}}
\label{evictionimpact_eval}
\end{figure}

In this experiment, we measure the performance impact on sender node when eviction happens in remote peer nodes(Figure \ref{evictionimpact_eval}). We set the same settings we used in Figure \ref{evictionimpactsetup}. Then, we run Redis with SYS workload because SYS workload has more write operations and this heavy write workload help us observe performance impact when remote eviction happens. After Redis populates peer nodes with about 17GB, we evict certain amount of victim MR blocks selected by Valet with activity-based victim selection. Then, we run Redis with YCSB SYS workload to measure the throughput. We repeat this up to 16GB eviction. Our observation indicates that Valet uses migration instead of eviction when remote eviction occurs and there is no performance impact on local node. However, without migration, one relies on batched-query-based random selection and remote eviction impact is significant on sender node. For example, 2GB eviction(about 8\% of workload) results in 50\% reduction of throughput on local node.

\section{Related Work}
\label{relatedwork}

\noindent
{\bf {Distributed Shared Memory/Disaggregated Memory.}\/} 
Although Distributed shared memory (DSM) was studied extensively \cite{amza1996treadmarks, bennett1990munin, li1989memory, Nelson+-usenixATC2015, scales1996shasta, schoinas1994fine}, DSM suffers poor performance due to high communication overhead. Disaggregated memory has attracted much attention recently and proposed new hardware architecture, and new network protocols to cut down the communication cost~\cite{asanovic2014firebox, faraboschi2015beyond, gao2016network, han2013network, lim2009disaggregated, rao2016memory}. Some proposals~\cite{mietke2006analysis, guo2016rdma, omni-path, tsai2017lite, zhu2015congestion, InfiniBand, recio2007remote, omni-path} show good ways to leverage RDMA technology by exploiting the disk-network latency gap. Remote storage for key-value stores\cite{dragojevic2014farm, dragojevic2015no, kalia2014using, mitchell2013using}, distributed objects \cite{waldo1996note}, object replication~\cite{Mojim} and swap pages~\cite{comer1990new, feeley1995implementing, FlourisMarkatos-JCC-1999, koussih1999dodo} show the benefit of RDMA technology in these use cases. Most of these efforts lack of desired transparency and all existing proposals treat and leverage unused host memory as the remote memory, fail to take advantage of the small performance gap between DRAM and Infiniband compared to disk. Effort to provide transparency at OS, network stack, or application level~\cite{Evangelos, Tia, Shuang, Haogan, Umesh, Juncheng, Hikari} has also been extensively studied. We put summary of comparion of these systems with Valet in Table \ref{systemcomparison}. However, these systems incur CPU overhead at receiver side, fail to handle remote eviction cost, lack of efficient local/remote resource orchestration or optimization in performance critical path.

\begin{table}[ht!]
\centering
\includegraphics[width=85mm]{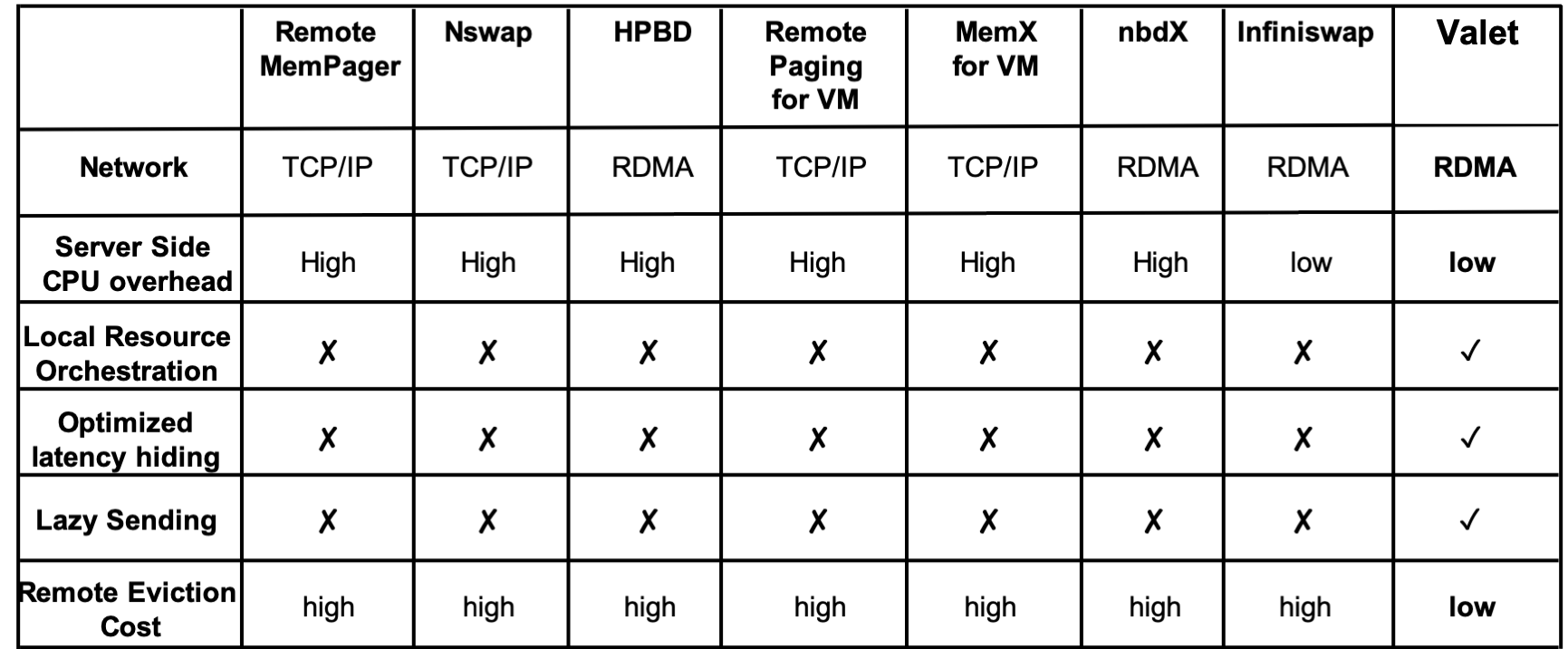}
\caption{Comparison with previous approaches.}
\label{systemcomparison}
\end{table}

\section{Conclusion}
\label{conclusion}

Valet addresses three common problems inherent in existing remote memory systems: latency overhead in the performance critical path, remote eviction impact and container-wide memory imbalance. We redesign the data flow in the critical path by introducing a host-coordinated memory pool that works as a local cache to reduce the latency in the critical path of the host and remote memory orchestration. Valet also tries to utilize unused local memory across containers by managing local memory via Valet host-coordinated memory pool, which allows containers to dynamically expand and shrink their memory allocations according to the workload demands. Valet provides an efficient remote memory reclaiming technique on remote nodes by an activity-based victim selection scheme to allow the least-active-block of data to be selected for serving the eviction requests and a migration protocol to move the least-active-block of data to less-memory-pressured remote node. Through extensive experiments on both big data and Machine Learning (ML) workloads, we show that Valet outperforms conventional OS swap by up to 138$\times$ and 282$\times$ for big data and ML workloads respectively, and by up to 5.3$\times$ completion improvement and by up to 3.5$\times$ completion improvement over the state-of-the-art remote paging systems for big data and ML workloads respectively. 

\subsection*{Acknowledgement} The first author thanks the opportunity of the 12-week working experience at IBM T. J. Watson Research Center in Summer 2019 with the group led by Donna N Dillenberger. This work is partially sponsored by the National Science Foundation under Grants NSF 2038029, NSF 2026945 and NSF 1564097, as well as an IBM faculty award.   

\bigskip
\bibliographystyle{plain}
\bibliography{\jobname}



\end{document}